\documentclass[%
 aip,
 amsmath,amssymb,
 reprint,%
]{revtex4-1}

\usepackage{graphicx}
\usepackage{dcolumn}
\usepackage{bm}

\usepackage[utf8]{inputenc}
\usepackage[T1]{fontenc}
\usepackage{mathptmx}
\usepackage{etoolbox}
\usepackage{upgreek} 
\usepackage{hyperref} 
\usepackage{multirow} 
\usepackage{newtxmath} 

\makeatletter
\def\@email#1#2{%
 \endgroup
 \patchcmd{\titleblock@produce}
  {\frontmatter@RRAPformat}
  {\frontmatter@RRAPformat{\produce@RRAP{*#1\href{mailto:#2}{#2}}}\frontmatter@RRAPformat}
  {}{}
}%
\makeatother
\begin{document}

\preprint{AIP/123-QED}

\title{ Enhanced Spectral Density of a Single Germanium Vacancy Center in a Nanodiamond by Cavity-Integration}   

\author{Florian Feuchtmayr}
\altaffiliation{F.F. and R.B.  contributed equally to this work.}
\affiliation{Institute for Quantum Optics, Ulm University, Albert-Einstein-Allee 11, 89081 Ulm, Germany} 
\author{Robert Berghaus}
\altaffiliation{F.F. and R.B.  contributed equally to this work.}
\affiliation{Institute for Quantum Optics, Ulm University, Albert-Einstein-Allee 11, 89081 Ulm, Germany}
\author{Selene Sachero}
\affiliation{Institute for Quantum Optics, Ulm University, Albert-Einstein-Allee 11, 89081 Ulm, Germany} 
\author{Gregor Bayer}
\affiliation{Institute for Quantum Optics, Ulm University, Albert-Einstein-Allee 11, 89081 Ulm, Germany} 
\author{Niklas Lettner}
\affiliation{Institute for Quantum Optics, Ulm University, Albert-Einstein-Allee 11, 89081 Ulm, Germany} 
\affiliation{Center for Integrated Quantum Science and Technology (IQst), Ulm University, Albert-Einstein-Allee 11, 89081 Ulm, Germany}
\author{Richard Waltrich}
\affiliation{Institute for Quantum Optics, Ulm University, Albert-Einstein-Allee 11, 89081 Ulm, Germany} 
\author{Patrick Maier}
\affiliation{Institute for Quantum Optics, Ulm University, Albert-Einstein-Allee 11, 89081 Ulm, Germany} 
\author{Viatcheslav Agafonov}
\affiliation{GREMAN, UMR 7347 CNRS, INSA-CVL, Tours University, 37200 Tours, France}
\author{Alexander Kubanek}
\altaffiliation[Corresponding author: ]{alexander.kubanek@uni-ulm.de}
\affiliation{Institute for Quantum Optics, Ulm University, Albert-Einstein-Allee 11, 89081 Ulm, Germany}
\affiliation{Center for Integrated Quantum Science and Technology (IQst), Ulm University, Albert-Einstein-Allee 11, 89081 Ulm, Germany}

\date{\today}

\begin{abstract} 
Color centers in diamond, among them the negatively-charged germanium vacancy (GeV$^-$), are promising candidates for many applications of quantum optics such as a quantum network. For efficient implementation, the optical transitions need to be coupled to a single optical mode. Here, we demonstrate the transfer of a nanodiamond containing a single ingrown GeV$^-$ center with excellent optical properties to an open Fabry-Pérot microcavity by nanomanipulation utilizing an atomic force microscope. Coupling of the GeV$^-$ defect to the cavity mode is achieved, while the optical resonator maintains a high finesse of $\mathcal{F}=7{,}700$ and a 48-fold spectral density enhancement is observed. This article demonstrates the integration of a GeV$^-$ defect with a Fabry-Pérot microcavity under ambient conditions with the potential to extend the experiments to cryogenic temperatures towards an efficient spin-photon platform.

\end{abstract}

\maketitle
\begin{flushleft}
\textit{The following article has been accepted by Applied Physics Letters. It is found at \href{https://doi.org/10.1063/5.0156787}{https://doi.org/10.1063/5.0156787}}
\end{flushleft}

With increasing advances in quantum technologies and the importance of secure communication, there is great interest in developing a quantum network. \cite{kimble2008quantum} In such a network, the communication is based on quantum entanglement, and by using quantum key distribution it provides guaranteed security. \cite{scarani2009security,Pirandola:20} 
However, to increase the range of the network on a global scale, \cite{Wehner2018QInternet} nodes are required due to lossy photon channels that limit the range to about 100\,km. \cite{jiang2009quantum}\\
Essential for the realization of such nodes in a scalable quantum network are a quantum memory capable of storing and processing quantum information, \cite{lim2006repeat} and high rates of coherent photons to establish entanglement in communication schemes. 
For the interaction between photons and the quantum memory, an efficient interface is necessary. Consequent challenges of overcoming photon losses, maintaining coherence, attaining robustness and  enabling scalability can be tackled by using a quantum repeater as a node in the quantum network. \cite{briegel1998} One approach to realizing such a repeater is to address a single emitter to gain access to a quantum memory via its spin. By placing the emitter inside a high quality optical resonator, \cite{vahala2003optical,Janitz:20}  quantum state writing and reading can be performed efficiently at high rates due to an enhanced spontaneous emission rate. \cite{purcell1946proceedings} \\
Promising single emitter candidates for realizing a quantum network are color centers, optically active defect centers in diamond. \cite{ruf2021quantum,chen2020building} In order to couple them to the mode of a microcavity diamond membranes (DMs) \cite{haussler2019diamond,Ruf2021Resonant} and  nanodiamonds (NDs) \cite{bayer_2022} are mainly used. Currently, negatively-charged nitrogen vacancies (NV$^-$) \cite{TheNV} and silicon vacancies (SiV$^-$) \cite{rogers2014} are the most studied color centers. While NV$^-$ centers show good spin properties and great strides have been made to entangle them, \cite{pompili2021} they show poor optical access, e.g. a low emission in the zero phonon line (ZPL). \cite{Jelezko2006Single} In contrast, group IV defects, with their interstitial doping atom in the diamond lattice, are promising alternative emitters for this concept. \cite{bradac2019quantum} Among them is the SiV$^-$, which compared to NV$^-$, has superior optical properties with a high Debye-Waller (DW) factor of 0.7. \cite{bradac2019quantum} A drawback is the technical overhead of mK temperatures, which is required for receiving good spin properties. \cite{lagomarsino2021creation}\\
The negatively-charged germanium vacancy (GeV$^-$) \cite{iwasaki2015germanium,nahra2021single} is another group IV defect with a heavier impurity atom than the SiV$^-$ center. 
The GeV$^-$ center exhibits a longer excited state lifetime and a comparable DW factor of 0.3-0.6. \cite{bradac2019quantum,Palyanov2015Germanium,nahra2021single,haussler2017photoluminescence}
At seven Kelvin, the phonon-induced spin dephasing is exponentially suppressed  resulting in good spin properties already below one Kelvin;  higher than  for the SiV$^-$ center. \cite{Janitz:20,Quantum2017Metsch}
Beside their implementation in diamond waveguides \cite{siampour2020ultrabright,Quantum2017Metsch} the GeV$^-$ defect was coupled to photonic resonators \cite{Bray2018,kumar2021fluorescence} and to one tunable Fabry-Pérot (FP) cavity. \cite{Hensen2020}
With a cavity formed by the end-facet of an optical fiber and a flat mirror with a bonded DM a spectral density enhancement (SDE) of 31 was reported. \cite{Hensen2020}\\
In this work, we demonstrate the transfer of a 200\,nm sized ND containing a single GeV$^-$ defect with non-blinking and narrow emission to an optical open FP microcavity with a high quality factor by utilizing the pick and place technique with an atomic force microscope (AFM). We realized coupling of the strong ZPL of the GeV$^-$ center to the cavity mode, while observing a 48-fold SDE and maintaining a high finesse. In our approach, the ND with the GeV$^-$ center is intrinsically aligned with the cavity field, making lateral scanning of the cavity mirror redundant. Therefore, the system is passively very stable with reduced technical overhead.\\
\begin{figure*}[!ht]    
    \centering
    \includegraphics[width=\textwidth]{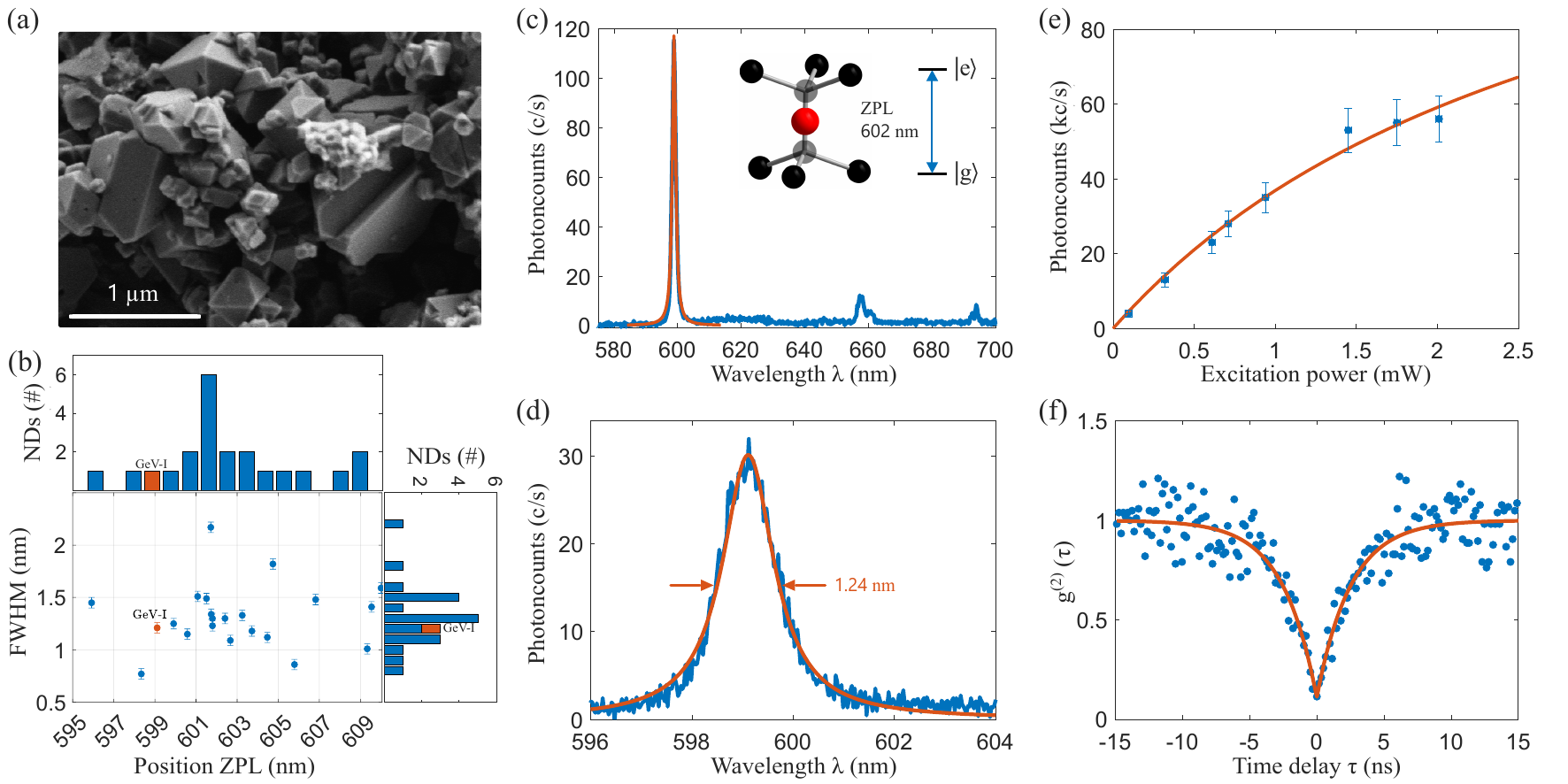}
    \caption{Characterizing GeV$^-$ center in NDs. (a) SEM image of the synthesized NDs. (b) Distribution of the ZPL and FWHM of the acquired GeV$^-$ defects. (c) PL spectrum with a coarse grating of 150\,g/mm revealing the ZPL and PSB of the selected ND GeV-I with inserted atomic structure of a GeV$^-$. (d) PL with a fine grating of 1200\,g/mm showing the ZPL of ND GeV-I at (599.11\,$\pm$\,0.03)\,nm and a FWHM of (1.24\,$\pm$\,0.03)\,nm under green excitation with (930\,$\pm$\,20)\,$\upmu$W. (e) Saturation curve of ND GeV-I with the saturation intensity $I_{\textup{sat}}$=(0.15\,$\pm$\,0.07)\,$\frac{\text{Mc}}{\text{s}}$ and saturation power $P_{\textup{sat}}$=(3.0\,$\pm$\,2.2)\,mW. (f) Off-resonant autocorrelation measurement without background correction of ND GeV-I showing strong anti-bunching of $g^{(2)}(0)$=0.11\,$\pm$\,0.04 and an excited state lifetime of $\tau_{\textup{LT}}$=(2.53\,$\pm$\,0.20)\,ns at (60\,$\pm$\,2)\,$\upmu$W.}
    \label{fig1}     
\end{figure*}
First, NDs with single GeV$^-$ defect centers were grown by high pressure high temperature (HPHT) \cite{palyanov2016high} synthesis under 1,450$^\circ$C and 8GPa. In this process, detonation NDs, fluoroadamantane (C$_{\text{10}}$H$_{\text{15}}$F) and hepta-fluoronaphtalene (C$_{\text{10}}$F$_{\text{8}}$) were used as carbon sources. Germanium triphenyl-chloride (GeC$_{\text{18}}$H$_{\text{15}}$Cl) was added as doping component for germanium. Fig. \ref{fig1}(a) shows a scanning electron microscope (SEM) image of the NDs of varying sizes after the synthesis. Subsequently, the NDs were cleaned by acid boiling and diluted in ultrapure water. After an ultrasonic bath for deagglomeration, the ND solution was drop-casted on a sapphire bulk substrate.\\
The sample was characterized in a confocal microscope (NA\,=\,0.9) at room temperature with an excitation wavelength of 532\,nm. NDs with emission from GeV$^-$ centers showed a strong ZPL at a median wavelength of (602\,$\pm$\,2)\,nm and are presented in fig. \ref{fig1}(b). The observed spectral distribution of the ZPL suggests strain in the NDs.\cite{Strain2018Meesala}
 With a mean FWHM of (1.3\,$\pm$\,0.3)\,nm  and a distribution between 0.8\,-\,2.2\,nm  the defects showed narrow linewidths compared to reported values. \cite{nahra2021single} 
The ND with the identifier GeV-I had good optical properties and was selected for further investigation. Its photoluminescence (PL) is displayed in figures \ref{fig1}(c,d). This color center demonstrated a high and stable emission rate in the ZPL with a high quality factor of $Q_{\textup{GeV}}=\lambda/\delta\lambda=483\,\pm\,12$. The spectrum shows a small phonon sideband (PSB) with a DW factor above 0.6, which results from different converging evaluations, see supplementary material. At 658\,nm and 693\,nm emission originated from an unrelated laser and the sapphire bulk, respectively, which was not taken into account for the DW factor. In addition, its saturation curve is illustrated in fig. \ref{fig1}(e) and was fitted with the function
\begin{equation}
   I(P)=\frac{I_{\textup{sat}}\times P}{P+P_{\textup{sat}}}.
\end{equation}
The determined values for saturation intensity and saturation power subside the presented data in Ref. \onlinecite{siampour2020ultrabright}. To evaluate the properties of the emission signal from the emitter, we performed an autocorrelation measurement under off-resonant excitation. The time resolved correlation without any background correction and with the fitting model 
\begin{equation}
    g^{(2)}(\tau)=1-A\times \exp\biggl(-\biggl|\frac{\tau}{\tau_{\textup{LT}}}\biggr|\biggr)
\end{equation}
 can be seen in fig. \ref{fig1}(f), for other emitters see supplementary material. A strong anti-bunching of $g^{(2)}(0)=0.11\,\pm\,0.04$ was observed. This shows a high rate of emission arising from only a single GeV$^-$ center within the ND under green excitation, making this a good candidate for quantum optics experiments and the use as a single photon source. An excited state lifetime of $\tau_{\textup{LT}}=(2.53\,\pm\,0.20)\,$ns at an excitation power of 60\,$\upmu$W was determined, in agreement with literature. \cite{iwasaki2015germanium}\\
\begin{figure*}[ht!]
    \centering
    \includegraphics[width=\textwidth]{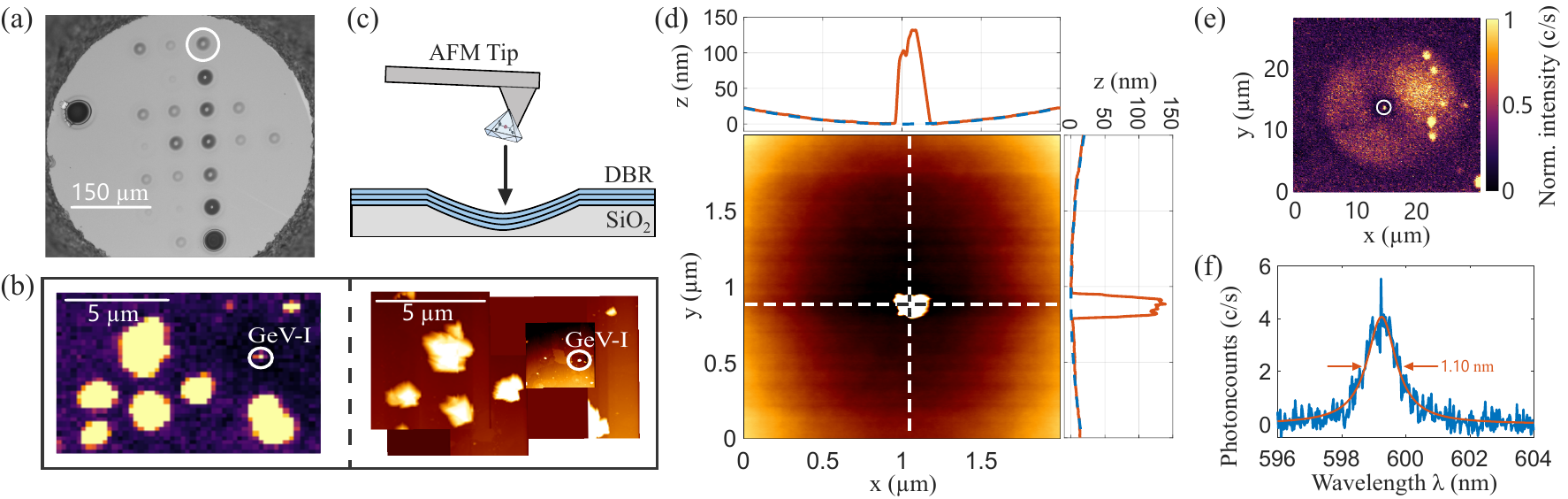}
    \caption{Assembly of an FP-resonator with a GeV$^-$ center containing ND. (a) Microscopic image of the mirror containing the curved structure (circled white) intended for the ND. (b) Confocal scan and AFM images of NDs showing GeV-I. (c) Schematic illustration of the pick and place technique for placing NDs inside a mirror structure using the AFM cantilever. (d) Tilt subtracted AFM image of the spherical mirror containing the transferred ND GeV-I in its center with respective cross sections in x- and y-directions. (e) Confocal scan image of the spherical mirror containing the same ND in its center after transfer. (f) PL of ND GeV-I in the mirror structure with a ZPL at (599.25\,$\pm$\,0.03)\,nm and a FWHM of (1.10\,$\pm$\,0.04)\,nm using an excitation power of (20\,$\pm$\,2)\,$\upmu$W.}
    \label{fig2}
\end{figure*}
Next, the GeV$^-$ center containing ND was integrated in the field of the hemispherical FP cavity by placing it in the center of the curved mirror. The concave mirror was fabricated by a CO$_2$ laser ablation process of a SiO$_2$ substrate. By creating multiple mirror structures on one substrate, a versatile mirror array could be fabricated. Varying the pulse lengths results in different structure sizes, as shown in fig. \ref{fig2}(a). Interferometric analysis of the structure intended for the ND revealed a depth of 2.7\,$\upmu$m and a diameter of 26.4\,$\upmu$m. By coating the structured substrate a distributed Bragg reflector (DBR) was formed, yielding a low transmission between 595-690nm with a minimum of $T\,<\,$310\,ppm at 601\,nm. At this wavelength the mirror coating  forms a field antinode around 150\,nm in air and 62\,nm in diamond away from the mirror surface.\\
Before the mirror integration, the emitter could be identified by comparing the confocal scan with the AFM image in fig. \ref{fig2}(b). Using the AFM based pick and place technique \cite{Fehler_2021,haussler2019preparing} nanomanipulation of ND GeV-I was achieved by transferring the emitter to the center of the desired structure, see fig. \ref{fig2}(c).\\
The mirror containing the ND was scanned with the AFM in fig. \ref{fig2}(d), whereby the cross sections reveal the size of the ND of (190\,x\,180\,x\,130)\,nm. It was also determined that the structure used here is elliptical, with radii of curvature RoC$_{\textup{X}}=(21.48\,\pm\,0.01)\,\upmu$m and RoC$_{\textup{Y}}=(28.94\,\pm\,0.01)\,\upmu$m. With the help of a confocal microscope, it could be confirmed that the optical properties were preserved. Fig. \ref{fig2}(e) shows the spherical shape of the mirror due to minor fluorescence of the coating. Pollution and other impurities are visible at the edge of the structure, outside the relevant cavity mode. In the center, a peak of fluorescence originating from ND GeV-I is observable, verified by the PL spectrum in fig. \ref{fig2}(f). The surrounding area of the ND shows a bleached background fluorescence due to previous laser scans with 532\,nm.\\
The spectral density (SD) of the spectrum 
\begin{equation}
    \textup{SD}=\frac{\textup{Photoncounts}}{\textup{FWHM}\times P}
\end{equation}
could be extracted by using the integrated photoncounts of the Lorentzian fit and the excitation power that lay within the linear range of the saturation curve. The spectrum in figure \ref{fig2}(f) yields a SD of $(12.7\,\pm\,1.0)$\,c/(s\,GHz\,mW), which could be compared to the free space emitter that demonstrates a SD of $(2.03\,\pm\,0.08)$\,c/(s\,GHz\,mW) in fig. \ref{fig1}(d). 
The measurements were performed on different confocal microscopes, so differences in the setup efficiencies had to be taken into account. With the reflected laser signal from the curved mirror and a different excitation beam width at the emitter, a relative excitation efficiency of $\Tilde{\eta}_{\textup{Con}}^{\textup{exc}}=\eta_{\textup{Con2}}^{\textup{exc}}/\eta_{\textup{Con1}}^{\textup{exc}}=2.3\,\pm\,0.2$ followed. Different optical components led to a relative detection efficiency of $\Tilde{\eta}_{\textup{Con}}^{\textup{det}}=1.3\,\pm\,0.2$. 
Considering these efficiencies and comparing the resulting SD from free space emission and the ND placed in the curved mirror suggested a spectral density enhancement of
$\textup{SDE}=\textup{SD}_{\textup{Con2}}/(\textup{SD}_{\textup{Con1}}\times\Tilde{\eta}_{\textup{Con}}^{\textup{det}}\times\Tilde{\eta}_{\textup{Con}}^{\textup{exc}})$$=2.1\,\pm\,0.4$, see supplementary material for complete calculations. 
This enhancement is attributed to the reflected fluorescence signal of the curved mirror, indicating a beneficial position of the emitter.
The 2-fold SDE suggesting efficient emitter coupling indicates that the emitter is located near the electric field maximum, which is estimated to be at a distance of 62 nm from the mirror surface.\\
\begin{figure}[tp]
    \centering
    \includegraphics[width=\linewidth]{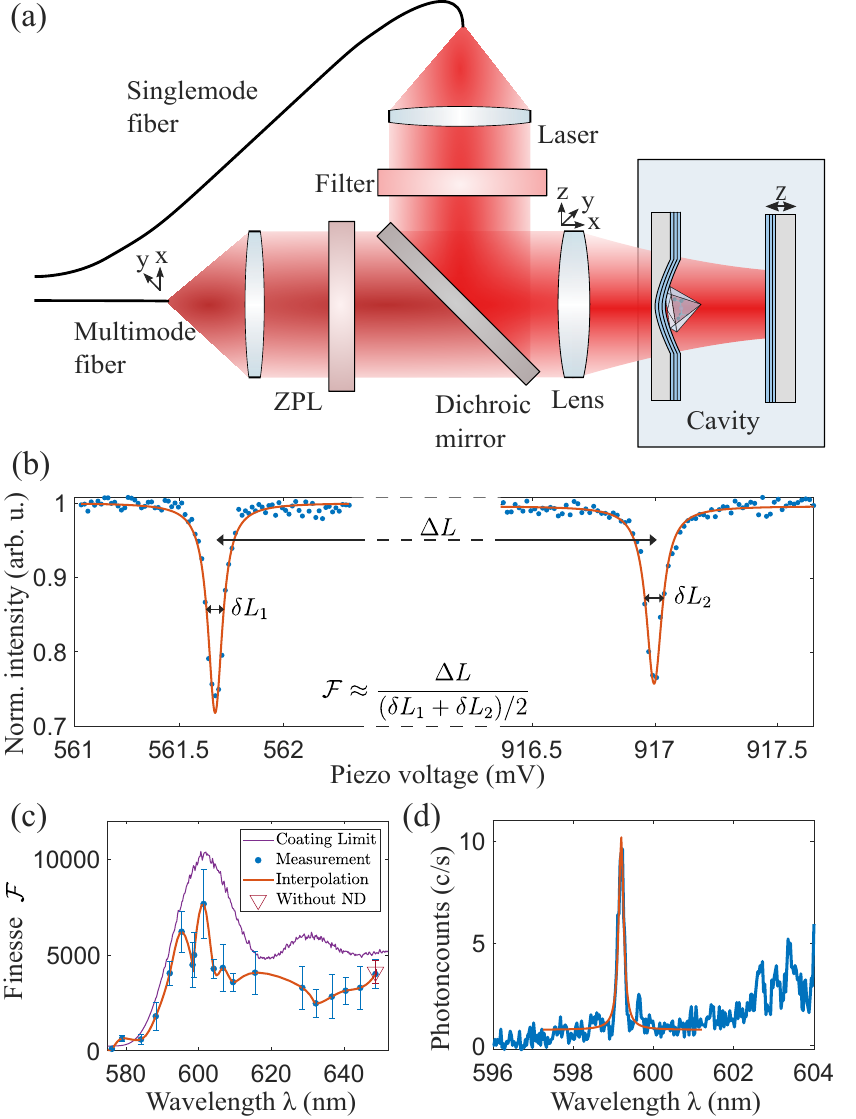}
    \caption{Characterizing the assembled cavity with a GeV$^-$ ND. 
    (a)~Schematic illustration of the operated optical microcavity setup. (b) Exemplary finesse measurement showing the reflected laser signal over the z piezo voltage of the flat mirror corresponding to the cavity length. (c) Possible finesse values of the microcavity determined by the transmission of the coating compared to the measurement with the ND inside the cavity. (d) PL of resonance n\,=\,51 of the optical microcavity with an excitation wavelength of 587.78\,nm and excitation power of (80\,$\pm$\,12)\,$\upmu$W. The spectrum (spectrometer resolution of $0.03$\,nm) shows a cavity modulated signal of the GeV$^-$ with a peak of emission at (599.20\,$\pm$\,0.03)\,nm. }
    \label{fig3}
\end{figure}
For characterizing the GeV$^-$ center in the cavity, the curved mirror containing the transferred ND and a second plane mirror with the same substrate and coating were assembled and formed the resonator, as illustrated in fig. \ref{fig3}(a). After passing a laser clean up filter (600 SP) and the reflection of a dichroic mirror, the excitation laser light (Sirah Matisse 2 DS Dye laser: linewidth below $100$\,kHz at 600\,nm) of a singlemode fiber (IVG Fiber: SM CU600PSC) was coupled into one of the cavity structures with the help of a lens (Asphericon: AHL10-08-P-U-780, \textit{f}\,=\,8mm, NA\,=\,0.55) mounted to nanopositioners in x, y and z (Attocube ANP series). An additional nanopositioner with z piezo enabled tunability of the cavity length by scanning the plane mirror. The cavity signal and the reflected light passed the lens, the dichroic mirror, a filter (605/15 BP) and were collected by a multimode fiber (Thorlabs: MM FG050LGA) in order to guide it to an APD or spectrometer.\\
The cavity finesse was extracted from the reflection signal by scanning the piezo of the plane mirror, as seen in fig. \ref{fig3}(b). The obtained values were compared to the theoretical possible finesse given by the coating limit in fig. \ref{fig3}(c). The cavity maintained a high finesse close to the coating limit. One could receive a maximum finesse of $\mathcal{F}>10{,}300$ at 601\,nm, while in the measurements a value of $\mathcal{F}=7{,}700\pm1{,}800$ was achieved. Comparing the finesse at 648.6\,nm before $\mathcal{F}=4{,}100\pm600$ and after $\mathcal{F}=4{,}000\pm800$ placing the ND in the cavity showed no significant losses due to reduced scattering of the small sized nanoparticle. \\
With an excitation wavelength of 587.8\,nm the cavity modulated ZPL signal of the single GeV$^-$ center was detected, as shown in  fig. \ref{fig3}(d). With the signal of the emitter at 599.2\,nm and the excitation wavelength at neighboring resonances of the resonator, a cavity length of $L=(15.72\,\pm\,0.06)\,\upmu$m was determined. Considering the RoC of $(25\,\pm\,5)\,\upmu$m, this resulted in a beam waist on the flat side of $\omega_0=(1.5\,\pm\,0.2)\,\upmu$m and on the curved side of $\omega(L)=(2.5\,\pm\,0.6)\,\upmu$m. From this a mode volume of $V=\frac{\pi}{4}L\omega_0^2=(140\,\pm\,40)\,\lambda_{\textup{Las}}^3$ followed and with the finesse of $\mathcal{F}=5{,}000\,\pm\,1{,}200$ at the emitter wavelength the quality factor of the cavity was evaluated to $Q_{\textup{Cav}}=\,260{,}000\,\pm\,60{,}000$. Finally, these parameters of the cavity enabled calculating the Purcell-factor with
\begin{equation}
    \mathcal{P}=\frac{3}{4\pi^2}\bigg(\frac{\lambda}{n}\bigg)^3\frac{Q_{\textup{Cav}}}{V},
\end{equation}
and the refraction index $n_{\textup{air}}=1$,
giving the values $\mathcal{P}_{\textup{flat}}=152\,\pm\,56$ on the flat side and $\mathcal{P}_{\textup{curv}}=57\,\pm\,29$ on the curved side.\\
However, this enhancement is only achieved for $Q_{\textup{Cav}}<Q_{\textup{GeV}}$, which was not the case for our room temperature experiments due to thermal broadening of the emitter.\cite{albrecht2013coupling} Therefore, the corrected formula for the Purcell effect is given by
\begin{equation}
    \mathcal{P}^*=\mathcal{P}\times\frac{Q_{\textup{GeV}}}{Q_{\textup{Cav}}},
\end{equation}
yielding the corrected values $\mathcal{P}_{\textup{flat}}^*=$\,0.28$\,\pm\,$0.03 on the flat side and $\mathcal{P}_{\textup{curv}}^*=$\,0.12$\,\pm\,$0.02 on the curved side.\\
In order to determine the SDE experimentally, we extracted the SD $(470\,\pm\,120)\,$c/(s\,GHz\,mW) of the emitter in the microcavity from fig. \ref{fig3}(d) and compared it to the free space emitter from fig. \ref{fig1}(d). We evaluated the altered excitation efficiency $\Tilde{\eta}_{\textup{Cav}}^{\textup{exc}}=11\,\pm\,4$ in relation to the confocal microscope setup, which was caused by the finesse of the cavity, a more efficient excitation wavelength \cite{haussler2017photoluminescence} and a wider excitation beam at the emitter. The detection efficiency was corrected by a factor of $\Tilde{\eta}_{\textup{Cav}}^{\textup{det}}=0.45\,\pm\,0.07$ due to different optical components. This resulted in a (48\,$\pm$\,20)-fold SDE, see  supplementary material for complete calculations. Considering that the Purcell effect is limited at room temperature, the enhancement can be attributed to cavity funneling.\\
In conclusion, we demonstrated the assembly of a tunable and versatile cavity system containing a single GeV$^-$ inside one ND. The ingrown GeV$^-$ defect center showed strong, stable and narrow emission in the ZPL with a small PSB yielding a DW of above 0.6. With a $g^{(2)}(0)$ value of 0.11\,$\pm$\,0.04 it proved to be a high purity single photon source. 
The small size of the ND enabled integration into the optical resonator mode without significant scattering losses, maintaining a high quality factor $Q_{\textup{Cav}}$ above $250{,}000$. The nanomanipulation capabilities allowed highly accurate alignment of the pre-characterized ND with GeV$^-$ center to the cavity field. 
With this integration into the open, fully tunable FP microcavity photoluminescence measurements could be performed under off-resonant excitation at room temperature, taking advantage of the single emitter emission.
We calculated a 48-fold SDE of the emitter and retained a high finesse of $\mathcal{F}=7{,}700$ including the nanoparticle.
 
The passively stable platform, with a mode volume V of $(140\,\pm\,40)\,\lambda_{\textup{Las}}^3$, does not require additional lateral scanning of the cavity mirror, profits from a low thermal expansion due to the main setup material choice of titanium, and allows for remote realignment with the nanopositioners. Altogether this offers robustness even at cryogenic temperatures. \cite{bayer_2022}
Cooling this platform to low temperatures, as shown in earlier work with SiV-centers, \cite{bayer_2022}  offers a decreased thermal broadening of the emitter towards the FT limit, which suggests an increased Purcell effect of $\mathcal{P}>50$. In combination with improved coherence of the emission into the ZPL, this results in a higher spectral enhancement of the signal out of the cavity.
In order to receive access to a quantum memory, spin access would be a desirable next goal to establish an efficient spin photon interface. With those advances this hybrid microcavity system is a robust and promising approach for many quantum optic applications such as a quantum repeater node.\\

\section*{Supplementary Material}

For additional information on the DW factor, additional autocorrelation measurements for other NDs, nanomanipulation of the ND, and comprehensive calculations of the spectral density enhancement, please refer to our supplementary material.

\begin{acknowledgments}
The authors thank V.A. Davydov for synthesis and processing of the ND material. 
The authors thank Jan Schimmel for the fabrication of the CO$_2$ structures and Lukas Antoniuk for experimental support. Funding by the German Federal Ministry of Education and Research (BMBF) within the project QR.X (16KISQ006) is gratefully acknowledged. S.S. acknowledges support of the Marie Curie ITN
project LasIonDef (GA n.956387). N.L. acknowledges support of IQst. 
Most measurements were performed on the basis of the Qudi software suite. \cite{BINDER2017}
\end{acknowledgments}

\section*{DATA AVAILABILITY}

The data that support the findings of this study are available
from the corresponding authors upon reasonable request.

\section*{AUTHOR DECLARATIONS}
\subsection*{Conflict of Interest}
The authors have no conflicts to disclose.

\subsection*{Author Contributions}
Florian Feuchtmayr and Robert Berghaus contributed equally to this work.\\
\textbf{Florian Feuchtmayr:} Investigation (equal); Data Curation (lead); Visualization (lead); Writing – original draft (lead); Writing – review and editing (equal). \textbf{Robert Berghaus:} Conceptualization (supporting); Methodology (equal); Investigation (equal); Data Curation (supporting); Visualization (supporting); Writing – original draft (supporting); Writing – review and editing (equal). \textbf{Selene Sachero:} Methodology (equal); Investigation (equal). \textbf{Gregor Bayer:} Conceptualization (supporting); Methodology (equal). \textbf{Niklas Lettner:} Investigation (equal). \textbf{Richard Waltrich:} Methodology (supporting). \textbf{Patrick Maier:} Methodology (supporting). \textbf{Viatcheslav Agafonov:} Methodology (equal). \textbf{Alexander Kubanek:} Conceptualization (lead); Supervision (lead); Funding acquisition (lead); Writing – review and editing (supporting).

\section*{References}

\bibliography{aipsamp}  

\providecommand{\noopsort}[1]{}\providecommand{\singleletter}[1]{#1}%
\begin{thebibliography}{36}%
\makeatletter
\providecommand \@ifxundefined [1]{%
 \@ifx{#1\undefined}
}%
\providecommand \@ifnum [1]{%
 \ifnum #1\expandafter \@firstoftwo
 \else \expandafter \@secondoftwo
 \fi
}%
\providecommand \@ifx [1]{%
 \ifx #1\expandafter \@firstoftwo
 \else \expandafter \@secondoftwo
 \fi
}%
\providecommand \natexlab [1]{#1}%
\providecommand \enquote  [1]{``#1''}%
\providecommand \bibnamefont  [1]{#1}%
\providecommand \bibfnamefont [1]{#1}%
\providecommand \citenamefont [1]{#1}%
\providecommand \href@noop [0]{\@secondoftwo}%
\providecommand \href [0]{\begingroup \@sanitize@url \@href}%
\providecommand \@href[1]{\@@startlink{#1}\@@href}%
\providecommand \@@href[1]{\endgroup#1\@@endlink}%
\providecommand \@sanitize@url [0]{\catcode `\\12\catcode `\$12\catcode
  `\&12\catcode `\#12\catcode `\^12\catcode `\_12\catcode `\%12\relax}%
\providecommand \@@startlink[1]{}%
\providecommand \@@endlink[0]{}%
\providecommand \url  [0]{\begingroup\@sanitize@url \@url }%
\providecommand \@url [1]{\endgroup\@href {#1}{\urlprefix }}%
\providecommand \urlprefix  [0]{URL }%
\providecommand \Eprint [0]{\href }%
\providecommand \doibase [0]{http://dx.doi.org/}%
\providecommand \selectlanguage [0]{\@gobble}%
\providecommand \bibinfo  [0]{\@secondoftwo}%
\providecommand \bibfield  [0]{\@secondoftwo}%
\providecommand \translation [1]{[#1]}%
\providecommand \BibitemOpen [0]{}%
\providecommand \bibitemStop [0]{}%
\providecommand \bibitemNoStop [0]{.\EOS\space}%
\providecommand \EOS [0]{\spacefactor3000\relax}%
\providecommand \BibitemShut  [1]{\csname bibitem#1\endcsname}%
\let\auto@bib@innerbib\@empty
\bibitem [{\citenamefont {Kimble}(2008)}]{kimble2008quantum}%
  \BibitemOpen
  \bibfield  {author} {\bibinfo {author} {\bibfnamefont {H.~J.}\ \bibnamefont
  {Kimble}},\ }\bibfield  {title} {\enquote {\bibinfo {title} {The quantum
  internet},}\ }\href {https://doi.org/10.1038/nature07127} {\bibfield
  {journal} {\bibinfo  {journal} {Nature}\ }\textbf {\bibinfo {volume} {453}},\
  \bibinfo {pages} {1023--1030} (\bibinfo {year} {2008})}\BibitemShut {NoStop}%
\bibitem [{\citenamefont {Scarani}\ \emph {et~al.}(2009)\citenamefont
  {Scarani}, \citenamefont {Bechmann-Pasquinucci}, \citenamefont {Cerf},
  \citenamefont {Du{\v{s}}ek}, \citenamefont {L{\"u}tkenhaus},\ and\
  \citenamefont {Peev}}]{scarani2009security}%
  \BibitemOpen
  \bibfield  {author} {\bibinfo {author} {\bibfnamefont {V.}~\bibnamefont
  {Scarani}}, \bibinfo {author} {\bibfnamefont {H.}~\bibnamefont
  {Bechmann-Pasquinucci}}, \bibinfo {author} {\bibfnamefont {N.~J.}\
  \bibnamefont {Cerf}}, \bibinfo {author} {\bibfnamefont {M.}~\bibnamefont
  {Du{\v{s}}ek}}, \bibinfo {author} {\bibfnamefont {N.}~\bibnamefont
  {L{\"u}tkenhaus}}, \ and\ \bibinfo {author} {\bibfnamefont {M.}~\bibnamefont
  {Peev}},\ }\bibfield  {title} {\enquote {\bibinfo {title} {The security of
  practical quantum key distribution},}\ }\href {\doibase
  10.1103/RevModPhys.81.1301} {\bibfield  {journal} {\bibinfo  {journal}
  {Reviews of modern physics}\ }\textbf {\bibinfo {volume} {81}},\ \bibinfo
  {pages} {1301} (\bibinfo {year} {2009})}\BibitemShut {NoStop}%
\bibitem [{\citenamefont {Pirandola}\ \emph {et~al.}(2020)\citenamefont
  {Pirandola}, \citenamefont {Andersen}, \citenamefont {Banchi}, \citenamefont
  {Berta}, \citenamefont {Bunandar}, \citenamefont {Colbeck}, \citenamefont
  {Englund}, \citenamefont {Gehring}, \citenamefont {Lupo}, \citenamefont
  {Ottaviani}, \citenamefont {Pereira}, \citenamefont {Razavi}, \citenamefont
  {Shaari}, \citenamefont {Tomamichel}, \citenamefont {Usenko}, \citenamefont
  {Vallone}, \citenamefont {Villoresi},\ and\ \citenamefont
  {Wallden}}]{Pirandola:20}%
  \BibitemOpen
  \bibfield  {author} {\bibinfo {author} {\bibfnamefont {S.}~\bibnamefont
  {Pirandola}}, \bibinfo {author} {\bibfnamefont {U.~L.}\ \bibnamefont
  {Andersen}}, \bibinfo {author} {\bibfnamefont {L.}~\bibnamefont {Banchi}},
  \bibinfo {author} {\bibfnamefont {M.}~\bibnamefont {Berta}}, \bibinfo
  {author} {\bibfnamefont {D.}~\bibnamefont {Bunandar}}, \bibinfo {author}
  {\bibfnamefont {R.}~\bibnamefont {Colbeck}}, \bibinfo {author} {\bibfnamefont
  {D.}~\bibnamefont {Englund}}, \bibinfo {author} {\bibfnamefont
  {T.}~\bibnamefont {Gehring}}, \bibinfo {author} {\bibfnamefont
  {C.}~\bibnamefont {Lupo}}, \bibinfo {author} {\bibfnamefont {C.}~\bibnamefont
  {Ottaviani}}, \bibinfo {author} {\bibfnamefont {J.~L.}\ \bibnamefont
  {Pereira}}, \bibinfo {author} {\bibfnamefont {M.}~\bibnamefont {Razavi}},
  \bibinfo {author} {\bibfnamefont {J.~S.}\ \bibnamefont {Shaari}}, \bibinfo
  {author} {\bibfnamefont {M.}~\bibnamefont {Tomamichel}}, \bibinfo {author}
  {\bibfnamefont {V.~C.}\ \bibnamefont {Usenko}}, \bibinfo {author}
  {\bibfnamefont {G.}~\bibnamefont {Vallone}}, \bibinfo {author} {\bibfnamefont
  {P.}~\bibnamefont {Villoresi}}, \ and\ \bibinfo {author} {\bibfnamefont
  {P.}~\bibnamefont {Wallden}},\ }\bibfield  {title} {\enquote {\bibinfo
  {title} {Advances in quantum cryptography},}\ }\href {\doibase
  10.1364/AOP.361502} {\bibfield  {journal} {\bibinfo  {journal} {Adv. Opt.
  Photon.}\ }\textbf {\bibinfo {volume} {12}},\ \bibinfo {pages} {1012--1236}
  (\bibinfo {year} {2020})}\BibitemShut {NoStop}%
\bibitem [{\citenamefont {Wehner}, \citenamefont {Elkouss},\ and\ \citenamefont
  {Hanson}(2018)}]{Wehner2018QInternet}%
  \BibitemOpen
  \bibfield  {author} {\bibinfo {author} {\bibfnamefont {S.}~\bibnamefont
  {Wehner}}, \bibinfo {author} {\bibfnamefont {D.}~\bibnamefont {Elkouss}}, \
  and\ \bibinfo {author} {\bibfnamefont {R.}~\bibnamefont {Hanson}},\
  }\bibfield  {title} {\enquote {\bibinfo {title} {Quantum internet: A vision
  for the road ahead},}\ }\href {\doibase 10.1126/science.aam9288} {\bibfield
  {journal} {\bibinfo  {journal} {Science}\ }\textbf {\bibinfo {volume}
  {362}},\ \bibinfo {pages} {eaam9288} (\bibinfo {year} {2018})}\BibitemShut
  {NoStop}%
\bibitem [{\citenamefont {Jiang}\ \emph {et~al.}(2009)\citenamefont {Jiang},
  \citenamefont {Taylor}, \citenamefont {Nemoto}, \citenamefont {Munro},
  \citenamefont {Van~Meter},\ and\ \citenamefont {Lukin}}]{jiang2009quantum}%
  \BibitemOpen
  \bibfield  {author} {\bibinfo {author} {\bibfnamefont {L.}~\bibnamefont
  {Jiang}}, \bibinfo {author} {\bibfnamefont {J.~M.}\ \bibnamefont {Taylor}},
  \bibinfo {author} {\bibfnamefont {K.}~\bibnamefont {Nemoto}}, \bibinfo
  {author} {\bibfnamefont {W.~J.}\ \bibnamefont {Munro}}, \bibinfo {author}
  {\bibfnamefont {R.}~\bibnamefont {Van~Meter}}, \ and\ \bibinfo {author}
  {\bibfnamefont {M.~D.}\ \bibnamefont {Lukin}},\ }\bibfield  {title} {\enquote
  {\bibinfo {title} {Quantum repeater with encoding},}\ }\href {\doibase
  10.1103/PhysRevA.79.032325} {\bibfield  {journal} {\bibinfo  {journal}
  {Physical Review A}\ }\textbf {\bibinfo {volume} {79}},\ \bibinfo {pages}
  {032325} (\bibinfo {year} {2009})}\BibitemShut {NoStop}%
\bibitem [{\citenamefont {Lim}\ \emph {et~al.}(2006)\citenamefont {Lim},
  \citenamefont {Barrett}, \citenamefont {Beige}, \citenamefont {Kok},\ and\
  \citenamefont {Kwek}}]{lim2006repeat}%
  \BibitemOpen
  \bibfield  {author} {\bibinfo {author} {\bibfnamefont {Y.~L.}\ \bibnamefont
  {Lim}}, \bibinfo {author} {\bibfnamefont {S.~D.}\ \bibnamefont {Barrett}},
  \bibinfo {author} {\bibfnamefont {A.}~\bibnamefont {Beige}}, \bibinfo
  {author} {\bibfnamefont {P.}~\bibnamefont {Kok}}, \ and\ \bibinfo {author}
  {\bibfnamefont {L.~C.}\ \bibnamefont {Kwek}},\ }\bibfield  {title} {\enquote
  {\bibinfo {title} {Repeat-until-success quantum computing using stationary
  and flying qubits},}\ }\href {\doibase 10.1103/PhysRevA.73.012304} {\bibfield
   {journal} {\bibinfo  {journal} {Physical Review A}\ }\textbf {\bibinfo
  {volume} {73}},\ \bibinfo {pages} {012304} (\bibinfo {year}
  {2006})}\BibitemShut {NoStop}%
\bibitem [{\citenamefont {Briegel}\ \emph {et~al.}(1998)\citenamefont
  {Briegel}, \citenamefont {D\"ur}, \citenamefont {Cirac},\ and\ \citenamefont
  {Zoller}}]{briegel1998}%
  \BibitemOpen
  \bibfield  {author} {\bibinfo {author} {\bibfnamefont {H.-J.}\ \bibnamefont
  {Briegel}}, \bibinfo {author} {\bibfnamefont {W.}~\bibnamefont {D\"ur}},
  \bibinfo {author} {\bibfnamefont {J.~I.}\ \bibnamefont {Cirac}}, \ and\
  \bibinfo {author} {\bibfnamefont {P.}~\bibnamefont {Zoller}},\ }\bibfield
  {title} {\enquote {\bibinfo {title} {Quantum repeaters: The role of imperfect
  local operations in quantum communication},}\ }\href {\doibase
  10.1103/PhysRevLett.81.5932} {\bibfield  {journal} {\bibinfo  {journal}
  {Phys. Rev. Lett.}\ }\textbf {\bibinfo {volume} {81}},\ \bibinfo {pages}
  {5932--5935} (\bibinfo {year} {1998})}\BibitemShut {NoStop}%
\bibitem [{\citenamefont {Vahala}(2003)}]{vahala2003optical}%
  \BibitemOpen
  \bibfield  {author} {\bibinfo {author} {\bibfnamefont {K.~J.}\ \bibnamefont
  {Vahala}},\ }\bibfield  {title} {\enquote {\bibinfo {title} {Optical
  microcavities},}\ }\href {https://doi.org/10.1038/nature01939} {\bibfield
  {journal} {\bibinfo  {journal} {nature}\ }\textbf {\bibinfo {volume} {424}},\
  \bibinfo {pages} {839--846} (\bibinfo {year} {2003})}\BibitemShut {NoStop}%
\bibitem [{\citenamefont {Janitz}, \citenamefont {Bhaskar},\ and\ \citenamefont
  {Childress}(2020)}]{Janitz:20}%
  \BibitemOpen
  \bibfield  {author} {\bibinfo {author} {\bibfnamefont {E.}~\bibnamefont
  {Janitz}}, \bibinfo {author} {\bibfnamefont {M.~K.}\ \bibnamefont {Bhaskar}},
  \ and\ \bibinfo {author} {\bibfnamefont {L.}~\bibnamefont {Childress}},\
  }\bibfield  {title} {\enquote {\bibinfo {title} {Cavity quantum
  electrodynamics with color centers in diamond},}\ }\href {\doibase
  10.1364/OPTICA.398628} {\bibfield  {journal} {\bibinfo  {journal} {Optica}\
  }\textbf {\bibinfo {volume} {7}},\ \bibinfo {pages} {1232--1252} (\bibinfo
  {year} {2020})}\BibitemShut {NoStop}%
\bibitem [{\citenamefont {Purcell}(1946)}]{purcell1946proceedings}%
  \BibitemOpen
  \bibfield  {author} {\bibinfo {author} {\bibfnamefont {E.~M.}\ \bibnamefont
  {Purcell}},\ }\bibfield  {title} {\enquote {\bibinfo {title} {Proceedings of
  the american physical society, b10. spontaneous emission probabilities at
  radio frequencies},}\ }\href@noop {} {\bibfield  {journal} {\bibinfo
  {journal} {Phys. Rev}\ }\textbf {\bibinfo {volume} {69}},\ \bibinfo {pages}
  {674} (\bibinfo {year} {1946})}\BibitemShut {NoStop}%
\bibitem [{\citenamefont {Ruf}\ \emph {et~al.}(2021{\natexlab{a}})\citenamefont
  {Ruf}, \citenamefont {Wan}, \citenamefont {Choi}, \citenamefont {Englund},\
  and\ \citenamefont {Hanson}}]{ruf2021quantum}%
  \BibitemOpen
  \bibfield  {author} {\bibinfo {author} {\bibfnamefont {M.}~\bibnamefont
  {Ruf}}, \bibinfo {author} {\bibfnamefont {N.~H.}\ \bibnamefont {Wan}},
  \bibinfo {author} {\bibfnamefont {H.}~\bibnamefont {Choi}}, \bibinfo {author}
  {\bibfnamefont {D.}~\bibnamefont {Englund}}, \ and\ \bibinfo {author}
  {\bibfnamefont {R.}~\bibnamefont {Hanson}},\ }\bibfield  {title} {\enquote
  {\bibinfo {title} {Quantum networks based on color centers in diamond},}\
  }\href {https://doi.org/10.1063/5.0056534} {\bibfield  {journal} {\bibinfo
  {journal} {Journal of Applied Physics}\ }\textbf {\bibinfo {volume} {130}},\
  \bibinfo {pages} {070901} (\bibinfo {year} {2021}{\natexlab{a}})}\BibitemShut
  {NoStop}%
\bibitem [{\citenamefont {Chen}, \citenamefont {Zheludev},\ and\ \citenamefont
  {Gao}(2020)}]{chen2020building}%
  \BibitemOpen
  \bibfield  {author} {\bibinfo {author} {\bibfnamefont {D.}~\bibnamefont
  {Chen}}, \bibinfo {author} {\bibfnamefont {N.}~\bibnamefont {Zheludev}}, \
  and\ \bibinfo {author} {\bibfnamefont {W.-b.}\ \bibnamefont {Gao}},\
  }\bibfield  {title} {\enquote {\bibinfo {title} {Building blocks for quantum
  network based on group-iv split-vacancy centers in diamond},}\ }\href
  {https://doi.org/10.1002/qute.201900069} {\bibfield  {journal} {\bibinfo
  {journal} {Advanced Quantum Technologies}\ }\textbf {\bibinfo {volume} {3}},\
  \bibinfo {pages} {1900069} (\bibinfo {year} {2020})}\BibitemShut {NoStop}%
\bibitem [{\citenamefont {H{\"a}u{\ss}ler}\ \emph
  {et~al.}(2019{\natexlab{a}})\citenamefont {H{\"a}u{\ss}ler}, \citenamefont
  {Benedikter}, \citenamefont {Bray}, \citenamefont {Regan}, \citenamefont
  {Dietrich}, \citenamefont {Twamley}, \citenamefont {Aharonovich},
  \citenamefont {Hunger},\ and\ \citenamefont {Kubanek}}]{haussler2019diamond}%
  \BibitemOpen
  \bibfield  {author} {\bibinfo {author} {\bibfnamefont {S.}~\bibnamefont
  {H{\"a}u{\ss}ler}}, \bibinfo {author} {\bibfnamefont {J.}~\bibnamefont
  {Benedikter}}, \bibinfo {author} {\bibfnamefont {K.}~\bibnamefont {Bray}},
  \bibinfo {author} {\bibfnamefont {B.}~\bibnamefont {Regan}}, \bibinfo
  {author} {\bibfnamefont {A.}~\bibnamefont {Dietrich}}, \bibinfo {author}
  {\bibfnamefont {J.}~\bibnamefont {Twamley}}, \bibinfo {author} {\bibfnamefont
  {I.}~\bibnamefont {Aharonovich}}, \bibinfo {author} {\bibfnamefont
  {D.}~\bibnamefont {Hunger}}, \ and\ \bibinfo {author} {\bibfnamefont
  {A.}~\bibnamefont {Kubanek}},\ }\bibfield  {title} {\enquote {\bibinfo
  {title} {Diamond photonics platform based on silicon vacancy centers in a
  single-crystal diamond membrane and a fiber cavity},}\ }\href {\doibase
  10.1103/PhysRevB.99.165310} {\bibfield  {journal} {\bibinfo  {journal}
  {Physical Review B}\ }\textbf {\bibinfo {volume} {99}},\ \bibinfo {pages}
  {165310} (\bibinfo {year} {2019}{\natexlab{a}})}\BibitemShut {NoStop}%
\bibitem [{\citenamefont {Ruf}\ \emph {et~al.}(2021{\natexlab{b}})\citenamefont
  {Ruf}, \citenamefont {Weaver}, \citenamefont {van Dam},\ and\ \citenamefont
  {Hanson}}]{Ruf2021Resonant}%
  \BibitemOpen
  \bibfield  {author} {\bibinfo {author} {\bibfnamefont {M.}~\bibnamefont
  {Ruf}}, \bibinfo {author} {\bibfnamefont {M.}~\bibnamefont {Weaver}},
  \bibinfo {author} {\bibfnamefont {S.}~\bibnamefont {van Dam}}, \ and\
  \bibinfo {author} {\bibfnamefont {R.}~\bibnamefont {Hanson}},\ }\bibfield
  {title} {\enquote {\bibinfo {title} {Resonant excitation and purcell
  enhancement of coherent nitrogen-vacancy centers coupled to a fabry-perot
  microcavity},}\ }\href {\doibase 10.1103/PhysRevApplied.15.024049} {\bibfield
   {journal} {\bibinfo  {journal} {Phys. Rev. Appl.}\ }\textbf {\bibinfo
  {volume} {15}},\ \bibinfo {pages} {024049} (\bibinfo {year}
  {2021}{\natexlab{b}})}\BibitemShut {NoStop}%
\bibitem [{\citenamefont {Bayer}\ \emph {et~al.}(2022)\citenamefont {Bayer},
  \citenamefont {Berghaus}, \citenamefont {Sachero}, \citenamefont
  {Filipovski}, \citenamefont {Antoniuk}, \citenamefont {Lettner},
  \citenamefont {Waltrich}, \citenamefont {Klotz}, \citenamefont {Maier},
  \citenamefont {Agafonov},\ and\ \citenamefont {Kubanek}}]{bayer_2022}%
  \BibitemOpen
  \bibfield  {author} {\bibinfo {author} {\bibfnamefont {G.}~\bibnamefont
  {Bayer}}, \bibinfo {author} {\bibfnamefont {R.}~\bibnamefont {Berghaus}},
  \bibinfo {author} {\bibfnamefont {S.}~\bibnamefont {Sachero}}, \bibinfo
  {author} {\bibfnamefont {A.~B.}\ \bibnamefont {Filipovski}}, \bibinfo
  {author} {\bibfnamefont {L.}~\bibnamefont {Antoniuk}}, \bibinfo {author}
  {\bibfnamefont {N.}~\bibnamefont {Lettner}}, \bibinfo {author} {\bibfnamefont
  {R.}~\bibnamefont {Waltrich}}, \bibinfo {author} {\bibfnamefont
  {M.}~\bibnamefont {Klotz}}, \bibinfo {author} {\bibfnamefont
  {P.}~\bibnamefont {Maier}}, \bibinfo {author} {\bibfnamefont
  {V.}~\bibnamefont {Agafonov}}, \ and\ \bibinfo {author} {\bibfnamefont
  {A.}~\bibnamefont {Kubanek}},\ }\bibfield  {title} {\enquote {\bibinfo
  {title} {A quantum repeater platform based on single siv$^-$ centers in
  diamond with cavity-assisted, all-optical spin access and fast coherent
  driving},}\ }\href {https://arxiv.org/abs/2210.16157} {\bibfield  {journal}
  {\bibinfo  {journal} {arXiv}\ } (\bibinfo {year} {2022})}\BibitemShut
  {NoStop}%
\bibitem [{\citenamefont {Doherty}\ \emph {et~al.}(2013)\citenamefont
  {Doherty}, \citenamefont {Manson}, \citenamefont {Delaney}, \citenamefont
  {Jelezko}, \citenamefont {Wrachtrup},\ and\ \citenamefont
  {Hollenberg}}]{TheNV}%
  \BibitemOpen
  \bibfield  {author} {\bibinfo {author} {\bibfnamefont {M.~W.}\ \bibnamefont
  {Doherty}}, \bibinfo {author} {\bibfnamefont {N.~B.}\ \bibnamefont {Manson}},
  \bibinfo {author} {\bibfnamefont {P.}~\bibnamefont {Delaney}}, \bibinfo
  {author} {\bibfnamefont {F.}~\bibnamefont {Jelezko}}, \bibinfo {author}
  {\bibfnamefont {J.}~\bibnamefont {Wrachtrup}}, \ and\ \bibinfo {author}
  {\bibfnamefont {L.~C.}\ \bibnamefont {Hollenberg}},\ }\bibfield  {title}
  {\enquote {\bibinfo {title} {The nitrogen-vacancy colour centre in
  diamond},}\ }\href {\doibase https://doi.org/10.1016/j.physrep.2013.02.001}
  {\bibfield  {journal} {\bibinfo  {journal} {Physics Reports}\ }\textbf
  {\bibinfo {volume} {528}},\ \bibinfo {pages} {1--45} (\bibinfo {year}
  {2013})}\BibitemShut {NoStop}%
\bibitem [{\citenamefont {Rogers}\ \emph {et~al.}(2014)\citenamefont {Rogers},
  \citenamefont {Jahnke}, \citenamefont {Doherty}, \citenamefont {Dietrich},
  \citenamefont {McGuinness}, \citenamefont {M\"uller}, \citenamefont {Teraji},
  \citenamefont {Sumiya}, \citenamefont {Isoya}, \citenamefont {Manson},\ and\
  \citenamefont {Jelezko}}]{rogers2014}%
  \BibitemOpen
  \bibfield  {author} {\bibinfo {author} {\bibfnamefont {L.~J.}\ \bibnamefont
  {Rogers}}, \bibinfo {author} {\bibfnamefont {K.~D.}\ \bibnamefont {Jahnke}},
  \bibinfo {author} {\bibfnamefont {M.~W.}\ \bibnamefont {Doherty}}, \bibinfo
  {author} {\bibfnamefont {A.}~\bibnamefont {Dietrich}}, \bibinfo {author}
  {\bibfnamefont {L.~P.}\ \bibnamefont {McGuinness}}, \bibinfo {author}
  {\bibfnamefont {C.}~\bibnamefont {M\"uller}}, \bibinfo {author}
  {\bibfnamefont {T.}~\bibnamefont {Teraji}}, \bibinfo {author} {\bibfnamefont
  {H.}~\bibnamefont {Sumiya}}, \bibinfo {author} {\bibfnamefont
  {J.}~\bibnamefont {Isoya}}, \bibinfo {author} {\bibfnamefont {N.~B.}\
  \bibnamefont {Manson}}, \ and\ \bibinfo {author} {\bibfnamefont
  {F.}~\bibnamefont {Jelezko}},\ }\bibfield  {title} {\enquote {\bibinfo
  {title} {Electronic structure of the negatively charged silicon-vacancy
  center in diamond},}\ }\href {\doibase 10.1103/PhysRevB.89.235101} {\bibfield
   {journal} {\bibinfo  {journal} {Phys. Rev. B}\ }\textbf {\bibinfo {volume}
  {89}},\ \bibinfo {pages} {235101} (\bibinfo {year} {2014})}\BibitemShut
  {NoStop}%
\bibitem [{\citenamefont {Pompili}\ \emph {et~al.}(2021)\citenamefont
  {Pompili}, \citenamefont {Hermans}, \citenamefont {Baier}, \citenamefont
  {Beukers}, \citenamefont {Humphreys}, \citenamefont {Schouten}, \citenamefont
  {Vermeulen}, \citenamefont {Tiggelman}, \citenamefont {dos Santos~Martins},
  \citenamefont {Dirkse}, \citenamefont {Wehner},\ and\ \citenamefont
  {Hanson}}]{pompili2021}%
  \BibitemOpen
  \bibfield  {author} {\bibinfo {author} {\bibfnamefont {M.}~\bibnamefont
  {Pompili}}, \bibinfo {author} {\bibfnamefont {S.~L.~N.}\ \bibnamefont
  {Hermans}}, \bibinfo {author} {\bibfnamefont {S.}~\bibnamefont {Baier}},
  \bibinfo {author} {\bibfnamefont {H.~K.~C.}\ \bibnamefont {Beukers}},
  \bibinfo {author} {\bibfnamefont {P.~C.}\ \bibnamefont {Humphreys}}, \bibinfo
  {author} {\bibfnamefont {R.~N.}\ \bibnamefont {Schouten}}, \bibinfo {author}
  {\bibfnamefont {R.~F.~L.}\ \bibnamefont {Vermeulen}}, \bibinfo {author}
  {\bibfnamefont {M.~J.}\ \bibnamefont {Tiggelman}}, \bibinfo {author}
  {\bibfnamefont {L.}~\bibnamefont {dos Santos~Martins}}, \bibinfo {author}
  {\bibfnamefont {B.}~\bibnamefont {Dirkse}}, \bibinfo {author} {\bibfnamefont
  {S.}~\bibnamefont {Wehner}}, \ and\ \bibinfo {author} {\bibfnamefont
  {R.}~\bibnamefont {Hanson}},\ }\bibfield  {title} {\enquote {\bibinfo {title}
  {Realization of a multinode quantum network of remote solid-state qubits},}\
  }\href {https://www.science.org/doi/abs/10.1126/science.abg1919} {\bibfield
  {journal} {\bibinfo  {journal} {Science}\ }\textbf {\bibinfo {volume}
  {372}},\ \bibinfo {pages} {259--264} (\bibinfo {year} {2021})}\BibitemShut
  {NoStop}%
\bibitem [{\citenamefont {Jelezko}\ and\ \citenamefont
  {Wrachtrup}(2006)}]{Jelezko2006Single}%
  \BibitemOpen
  \bibfield  {author} {\bibinfo {author} {\bibfnamefont {F.}~\bibnamefont
  {Jelezko}}\ and\ \bibinfo {author} {\bibfnamefont {J.}~\bibnamefont
  {Wrachtrup}},\ }\bibfield  {title} {\enquote {\bibinfo {title} {Single defect
  centres in diamond: A review},}\ }\href {\doibase
  https://doi.org/10.1002/pssa.200671403} {\bibfield  {journal} {\bibinfo
  {journal} {physica status solidi (a)}\ }\textbf {\bibinfo {volume} {203}},\
  \bibinfo {pages} {3207--3225} (\bibinfo {year} {2006})}\BibitemShut {NoStop}%
\bibitem [{\citenamefont {Bradac}\ \emph {et~al.}(2019)\citenamefont {Bradac},
  \citenamefont {Gao}, \citenamefont {Forneris}, \citenamefont {Trusheim},\
  and\ \citenamefont {Aharonovich}}]{bradac2019quantum}%
  \BibitemOpen
  \bibfield  {author} {\bibinfo {author} {\bibfnamefont {C.}~\bibnamefont
  {Bradac}}, \bibinfo {author} {\bibfnamefont {W.}~\bibnamefont {Gao}},
  \bibinfo {author} {\bibfnamefont {J.}~\bibnamefont {Forneris}}, \bibinfo
  {author} {\bibfnamefont {M.~E.}\ \bibnamefont {Trusheim}}, \ and\ \bibinfo
  {author} {\bibfnamefont {I.}~\bibnamefont {Aharonovich}},\ }\bibfield
  {title} {\enquote {\bibinfo {title} {Quantum nanophotonics with group iv
  defects in diamond},}\ }\href {https://doi.org/10.1038/s41467-019-13332-w}
  {\bibfield  {journal} {\bibinfo  {journal} {Nature communications}\ }\textbf
  {\bibinfo {volume} {10}},\ \bibinfo {pages} {1--13} (\bibinfo {year}
  {2019})}\BibitemShut {NoStop}%
\bibitem [{\citenamefont {Lagomarsino}\ \emph {et~al.}(2021)\citenamefont
  {Lagomarsino}, \citenamefont {Flatae}, \citenamefont {Kambalathmana},
  \citenamefont {Sledz}, \citenamefont {Hunold}, \citenamefont {Soltani},
  \citenamefont {Reuschel}, \citenamefont {Sciortino}, \citenamefont {Gelli},
  \citenamefont {Massi} \emph {et~al.}}]{lagomarsino2021creation}%
  \BibitemOpen
  \bibfield  {author} {\bibinfo {author} {\bibfnamefont {S.}~\bibnamefont
  {Lagomarsino}}, \bibinfo {author} {\bibfnamefont {A.}~\bibnamefont {Flatae}},
  \bibinfo {author} {\bibfnamefont {H.}~\bibnamefont {Kambalathmana}}, \bibinfo
  {author} {\bibfnamefont {F.}~\bibnamefont {Sledz}}, \bibinfo {author}
  {\bibfnamefont {L.}~\bibnamefont {Hunold}}, \bibinfo {author} {\bibfnamefont
  {N.}~\bibnamefont {Soltani}}, \bibinfo {author} {\bibfnamefont
  {P.}~\bibnamefont {Reuschel}}, \bibinfo {author} {\bibfnamefont
  {S.}~\bibnamefont {Sciortino}}, \bibinfo {author} {\bibfnamefont
  {N.}~\bibnamefont {Gelli}}, \bibinfo {author} {\bibfnamefont
  {M.}~\bibnamefont {Massi}},  \emph {et~al.},\ }\bibfield  {title} {\enquote
  {\bibinfo {title} {Creation of silicon-vacancy color centers in diamond by
  ion implantation},}\ }\href {https://doi.org/10.3389/fphy.2020.601362}
  {\bibfield  {journal} {\bibinfo  {journal} {Frontiers in Physics}\ }\textbf
  {\bibinfo {volume} {8}},\ \bibinfo {pages} {601362} (\bibinfo {year}
  {2021})}\BibitemShut {NoStop}%
\bibitem [{\citenamefont {Iwasaki}\ \emph {et~al.}(2015)\citenamefont
  {Iwasaki}, \citenamefont {Ishibashi}, \citenamefont {Miyamoto}, \citenamefont
  {Kobayashi}, \citenamefont {Miyazaki}, \citenamefont {Tahara}, \citenamefont
  {Jahnke}, \citenamefont {Rogers}, \citenamefont {Naydenov}, \citenamefont
  {Jelezko} \emph {et~al.}}]{iwasaki2015germanium}%
  \BibitemOpen
  \bibfield  {author} {\bibinfo {author} {\bibfnamefont {T.}~\bibnamefont
  {Iwasaki}}, \bibinfo {author} {\bibfnamefont {F.}~\bibnamefont {Ishibashi}},
  \bibinfo {author} {\bibfnamefont {Y.}~\bibnamefont {Miyamoto}}, \bibinfo
  {author} {\bibfnamefont {S.}~\bibnamefont {Kobayashi}}, \bibinfo {author}
  {\bibfnamefont {T.}~\bibnamefont {Miyazaki}}, \bibinfo {author}
  {\bibfnamefont {K.}~\bibnamefont {Tahara}}, \bibinfo {author} {\bibfnamefont
  {K.~D.}\ \bibnamefont {Jahnke}}, \bibinfo {author} {\bibfnamefont {L.~J.}\
  \bibnamefont {Rogers}}, \bibinfo {author} {\bibfnamefont {B.}~\bibnamefont
  {Naydenov}}, \bibinfo {author} {\bibfnamefont {F.}~\bibnamefont {Jelezko}},
  \emph {et~al.},\ }\bibfield  {title} {\enquote {\bibinfo {title}
  {Germanium-vacancy single color centers in diamond},}\ }\href
  {https://doi.org/10.1038/srep12882} {\bibfield  {journal} {\bibinfo
  {journal} {Scientific reports}\ }\textbf {\bibinfo {volume} {5}},\ \bibinfo
  {pages} {1--7} (\bibinfo {year} {2015})}\BibitemShut {NoStop}%
\bibitem [{\citenamefont {Nahra}\ \emph {et~al.}(2021)\citenamefont {Nahra},
  \citenamefont {Alshamaa}, \citenamefont {Deturche}, \citenamefont {Davydov},
  \citenamefont {Kulikova}, \citenamefont {Agafonov},\ and\ \citenamefont
  {Couteau}}]{nahra2021single}%
  \BibitemOpen
  \bibfield  {author} {\bibinfo {author} {\bibfnamefont {M.}~\bibnamefont
  {Nahra}}, \bibinfo {author} {\bibfnamefont {D.}~\bibnamefont {Alshamaa}},
  \bibinfo {author} {\bibfnamefont {R.}~\bibnamefont {Deturche}}, \bibinfo
  {author} {\bibfnamefont {V.}~\bibnamefont {Davydov}}, \bibinfo {author}
  {\bibfnamefont {L.}~\bibnamefont {Kulikova}}, \bibinfo {author}
  {\bibfnamefont {V.}~\bibnamefont {Agafonov}}, \ and\ \bibinfo {author}
  {\bibfnamefont {C.}~\bibnamefont {Couteau}},\ }\bibfield  {title} {\enquote
  {\bibinfo {title} {Single germanium vacancy centers in nanodiamonds with
  bulk-like spectral stability},}\ }\href {https://doi.org/10.1116/5.0035937}
  {\bibfield  {journal} {\bibinfo  {journal} {AVS Quantum Science}\ }\textbf
  {\bibinfo {volume} {3}} (\bibinfo {year} {2021})}\BibitemShut {NoStop}%
\bibitem [{\citenamefont {Palyanov}\ \emph {et~al.}(2015)\citenamefont
  {Palyanov}, \citenamefont {Kupriyanov}, \citenamefont {Borzdov},\ and\
  \citenamefont {Surovtsev}}]{Palyanov2015Germanium}%
  \BibitemOpen
  \bibfield  {author} {\bibinfo {author} {\bibfnamefont {Y.~N.}\ \bibnamefont
  {Palyanov}}, \bibinfo {author} {\bibfnamefont {I.~N.}\ \bibnamefont
  {Kupriyanov}}, \bibinfo {author} {\bibfnamefont {Y.~M.}\ \bibnamefont
  {Borzdov}}, \ and\ \bibinfo {author} {\bibfnamefont {N.~V.}\ \bibnamefont
  {Surovtsev}},\ }\bibfield  {title} {\enquote {\bibinfo {title} {Germanium: a
  new catalyst for diamond synthesis and a new optically active impurity in
  diamond.}}\ }\href {https://doi.org/10.1038/srep14789} {\bibfield  {journal}
  {\bibinfo  {journal} {Sci Rep}\ }\textbf {\bibinfo {volume} {5}},\ \bibinfo
  {pages} {14789} (\bibinfo {year} {2015})}\BibitemShut {NoStop}%
\bibitem [{\citenamefont {H{\"a}u{\ss}ler}\ \emph {et~al.}(2017)\citenamefont
  {H{\"a}u{\ss}ler}, \citenamefont {Thiering}, \citenamefont {Dietrich},
  \citenamefont {Waasem}, \citenamefont {Teraji}, \citenamefont {Isoya},
  \citenamefont {Iwasaki}, \citenamefont {Hatano}, \citenamefont {Jelezko},
  \citenamefont {Gali} \emph {et~al.}}]{haussler2017photoluminescence}%
  \BibitemOpen
  \bibfield  {author} {\bibinfo {author} {\bibfnamefont {S.}~\bibnamefont
  {H{\"a}u{\ss}ler}}, \bibinfo {author} {\bibfnamefont {G.}~\bibnamefont
  {Thiering}}, \bibinfo {author} {\bibfnamefont {A.}~\bibnamefont {Dietrich}},
  \bibinfo {author} {\bibfnamefont {N.}~\bibnamefont {Waasem}}, \bibinfo
  {author} {\bibfnamefont {T.}~\bibnamefont {Teraji}}, \bibinfo {author}
  {\bibfnamefont {J.}~\bibnamefont {Isoya}}, \bibinfo {author} {\bibfnamefont
  {T.}~\bibnamefont {Iwasaki}}, \bibinfo {author} {\bibfnamefont
  {M.}~\bibnamefont {Hatano}}, \bibinfo {author} {\bibfnamefont
  {F.}~\bibnamefont {Jelezko}}, \bibinfo {author} {\bibfnamefont
  {A.}~\bibnamefont {Gali}},  \emph {et~al.},\ }\bibfield  {title} {\enquote
  {\bibinfo {title} {Photoluminescence excitation spectroscopy of siv- and gev-
  color center in diamond},}\ }\href {\doibase 10.1088/1367-2630/aa73e5}
  {\bibfield  {journal} {\bibinfo  {journal} {New Journal of Physics}\ }\textbf
  {\bibinfo {volume} {19}},\ \bibinfo {pages} {063036} (\bibinfo {year}
  {2017})}\BibitemShut {NoStop}%
\bibitem [{\citenamefont {Bhaskar}\ \emph {et~al.}(2017)\citenamefont
  {Bhaskar}, \citenamefont {Sukachev}, \citenamefont {Sipahigil}, \citenamefont
  {Evans}, \citenamefont {Burek}, \citenamefont {Nguyen}, \citenamefont
  {Rogers}, \citenamefont {Siyushev}, \citenamefont {Metsch}, \citenamefont
  {Park}, \citenamefont {Jelezko}, \citenamefont {Lon\ifmmode~\check{c}\else
  \v{c}\fi{}ar},\ and\ \citenamefont {Lukin}}]{Quantum2017Metsch}%
  \BibitemOpen
  \bibfield  {author} {\bibinfo {author} {\bibfnamefont {M.~K.}\ \bibnamefont
  {Bhaskar}}, \bibinfo {author} {\bibfnamefont {D.~D.}\ \bibnamefont
  {Sukachev}}, \bibinfo {author} {\bibfnamefont {A.}~\bibnamefont {Sipahigil}},
  \bibinfo {author} {\bibfnamefont {R.~E.}\ \bibnamefont {Evans}}, \bibinfo
  {author} {\bibfnamefont {M.~J.}\ \bibnamefont {Burek}}, \bibinfo {author}
  {\bibfnamefont {C.~T.}\ \bibnamefont {Nguyen}}, \bibinfo {author}
  {\bibfnamefont {L.~J.}\ \bibnamefont {Rogers}}, \bibinfo {author}
  {\bibfnamefont {P.}~\bibnamefont {Siyushev}}, \bibinfo {author}
  {\bibfnamefont {M.~H.}\ \bibnamefont {Metsch}}, \bibinfo {author}
  {\bibfnamefont {H.}~\bibnamefont {Park}}, \bibinfo {author} {\bibfnamefont
  {F.}~\bibnamefont {Jelezko}}, \bibinfo {author} {\bibfnamefont
  {M.}~\bibnamefont {Lon\ifmmode~\check{c}\else \v{c}\fi{}ar}}, \ and\ \bibinfo
  {author} {\bibfnamefont {M.~D.}\ \bibnamefont {Lukin}},\ }\bibfield  {title}
  {\enquote {\bibinfo {title} {Quantum nonlinear optics with a
  germanium-vacancy color center in a nanoscale diamond waveguide},}\ }\href
  {\doibase 10.1103/PhysRevLett.118.223603} {\bibfield  {journal} {\bibinfo
  {journal} {Phys. Rev. Lett.}\ }\textbf {\bibinfo {volume} {118}},\ \bibinfo
  {pages} {223603} (\bibinfo {year} {2017})}\BibitemShut {NoStop}%
\bibitem [{\citenamefont {Siampour}\ \emph {et~al.}(2020)\citenamefont
  {Siampour}, \citenamefont {Wang}, \citenamefont {Zenin}, \citenamefont
  {Boroviks}, \citenamefont {Siyushev}, \citenamefont {Yang}, \citenamefont
  {Davydov}, \citenamefont {Kulikova}, \citenamefont {Agafonov}, \citenamefont
  {Kubanek}, \citenamefont {Mortensen}, \citenamefont {Jelezko},\ and\
  \citenamefont {Bozhevolnyi}}]{siampour2020ultrabright}%
  \BibitemOpen
  \bibfield  {author} {\bibinfo {author} {\bibfnamefont {H.}~\bibnamefont
  {Siampour}}, \bibinfo {author} {\bibfnamefont {O.}~\bibnamefont {Wang}},
  \bibinfo {author} {\bibfnamefont {V.~A.}\ \bibnamefont {Zenin}}, \bibinfo
  {author} {\bibfnamefont {S.}~\bibnamefont {Boroviks}}, \bibinfo {author}
  {\bibfnamefont {P.}~\bibnamefont {Siyushev}}, \bibinfo {author}
  {\bibfnamefont {Y.}~\bibnamefont {Yang}}, \bibinfo {author} {\bibfnamefont
  {V.~A.}\ \bibnamefont {Davydov}}, \bibinfo {author} {\bibfnamefont {L.~F.}\
  \bibnamefont {Kulikova}}, \bibinfo {author} {\bibfnamefont {V.~N.}\
  \bibnamefont {Agafonov}}, \bibinfo {author} {\bibfnamefont {A.}~\bibnamefont
  {Kubanek}}, \bibinfo {author} {\bibfnamefont {N.~A.}\ \bibnamefont
  {Mortensen}}, \bibinfo {author} {\bibfnamefont {F.}~\bibnamefont {Jelezko}},
  \ and\ \bibinfo {author} {\bibfnamefont {S.~I.}\ \bibnamefont
  {Bozhevolnyi}},\ }\bibfield  {title} {\enquote {\bibinfo {title} {Ultrabright
  single-photon emission from germanium-vacancy zero-phonon lines:
  deterministic emitter-waveguide interfacing at plasmonic hot spots},}\ }\href
  {\doibase doi:10.1515/nanoph-2020-0036} {\bibfield  {journal} {\bibinfo
  {journal} {Nanophotonics}\ }\textbf {\bibinfo {volume} {9}},\ \bibinfo
  {pages} {953--962} (\bibinfo {year} {2020})}\BibitemShut {NoStop}%
\bibitem [{\citenamefont {Bray}\ \emph {et~al.}(2018)\citenamefont {Bray},
  \citenamefont {Regan}, \citenamefont {Trycz}, \citenamefont {Previdi},
  \citenamefont {Seniutinas}, \citenamefont {Ganesan}, \citenamefont
  {Kianinia}, \citenamefont {Kim},\ and\ \citenamefont
  {Aharonovich}}]{Bray2018}%
  \BibitemOpen
  \bibfield  {author} {\bibinfo {author} {\bibfnamefont {K.}~\bibnamefont
  {Bray}}, \bibinfo {author} {\bibfnamefont {B.}~\bibnamefont {Regan}},
  \bibinfo {author} {\bibfnamefont {A.}~\bibnamefont {Trycz}}, \bibinfo
  {author} {\bibfnamefont {R.}~\bibnamefont {Previdi}}, \bibinfo {author}
  {\bibfnamefont {G.}~\bibnamefont {Seniutinas}}, \bibinfo {author}
  {\bibfnamefont {K.}~\bibnamefont {Ganesan}}, \bibinfo {author} {\bibfnamefont
  {M.}~\bibnamefont {Kianinia}}, \bibinfo {author} {\bibfnamefont
  {S.}~\bibnamefont {Kim}}, \ and\ \bibinfo {author} {\bibfnamefont
  {I.}~\bibnamefont {Aharonovich}},\ }\bibfield  {title} {\enquote {\bibinfo
  {title} {Single crystal diamond membranes and photonic resonators containing
  germanium vacancy color centers},}\ }\href
  {https://doi.org/10.1021/acsphotonics.8b00930} {\bibfield  {journal}
  {\bibinfo  {journal} {ACS Photonics}\ }\textbf {\bibinfo {volume} {5}},\
  \bibinfo {pages} {4817--4822} (\bibinfo {year} {2018})}\BibitemShut {NoStop}%
\bibitem [{\citenamefont {Kumar}\ \emph {et~al.}(2021)\citenamefont {Kumar},
  \citenamefont {Wu}, \citenamefont {Komisar}, \citenamefont {Kan},
  \citenamefont {Kulikova}, \citenamefont {Davydov}, \citenamefont {Agafonov},\
  and\ \citenamefont {Bozhevolnyi}}]{kumar2021fluorescence}%
  \BibitemOpen
  \bibfield  {author} {\bibinfo {author} {\bibfnamefont {S.}~\bibnamefont
  {Kumar}}, \bibinfo {author} {\bibfnamefont {C.}~\bibnamefont {Wu}}, \bibinfo
  {author} {\bibfnamefont {D.}~\bibnamefont {Komisar}}, \bibinfo {author}
  {\bibfnamefont {Y.}~\bibnamefont {Kan}}, \bibinfo {author} {\bibfnamefont
  {L.~F.}\ \bibnamefont {Kulikova}}, \bibinfo {author} {\bibfnamefont {V.~A.}\
  \bibnamefont {Davydov}}, \bibinfo {author} {\bibfnamefont {V.~N.}\
  \bibnamefont {Agafonov}}, \ and\ \bibinfo {author} {\bibfnamefont {S.~I.}\
  \bibnamefont {Bozhevolnyi}},\ }\bibfield  {title} {\enquote {\bibinfo {title}
  {Fluorescence enhancement of a single germanium vacancy center in a
  nanodiamond by a plasmonic bragg cavity},}\ }\href
  {https://doi.org/10.1063/5.0033507} {\bibfield  {journal} {\bibinfo
  {journal} {The Journal of Chemical Physics}\ }\textbf {\bibinfo {volume}
  {154}},\ \bibinfo {pages} {044303} (\bibinfo {year} {2021})}\BibitemShut
  {NoStop}%
\bibitem [{\citenamefont {H\o{}y~Jensen}\ \emph {et~al.}(2020)\citenamefont
  {H\o{}y~Jensen}, \citenamefont {Janitz}, \citenamefont {Fontana},
  \citenamefont {He}, \citenamefont {Gobron}, \citenamefont {Radko},
  \citenamefont {Bhaskar}, \citenamefont {Evans}, \citenamefont
  {Rodr\'{\i}guez~Rosenblueth}, \citenamefont {Childress}, \citenamefont
  {Huck},\ and\ \citenamefont {Lund~Andersen}}]{Hensen2020}%
  \BibitemOpen
  \bibfield  {author} {\bibinfo {author} {\bibfnamefont {R.}~\bibnamefont
  {H\o{}y~Jensen}}, \bibinfo {author} {\bibfnamefont {E.}~\bibnamefont
  {Janitz}}, \bibinfo {author} {\bibfnamefont {Y.}~\bibnamefont {Fontana}},
  \bibinfo {author} {\bibfnamefont {Y.}~\bibnamefont {He}}, \bibinfo {author}
  {\bibfnamefont {O.}~\bibnamefont {Gobron}}, \bibinfo {author} {\bibfnamefont
  {I.~P.}\ \bibnamefont {Radko}}, \bibinfo {author} {\bibfnamefont
  {M.}~\bibnamefont {Bhaskar}}, \bibinfo {author} {\bibfnamefont
  {R.}~\bibnamefont {Evans}}, \bibinfo {author} {\bibfnamefont {C.~D.}\
  \bibnamefont {Rodr\'{\i}guez~Rosenblueth}}, \bibinfo {author} {\bibfnamefont
  {L.}~\bibnamefont {Childress}}, \bibinfo {author} {\bibfnamefont
  {A.}~\bibnamefont {Huck}}, \ and\ \bibinfo {author} {\bibfnamefont
  {U.}~\bibnamefont {Lund~Andersen}},\ }\bibfield  {title} {\enquote {\bibinfo
  {title} {Cavity-enhanced photon emission from a single germanium-vacancy
  center in a diamond membrane},}\ }\href {\doibase
  10.1103/PhysRevApplied.13.064016} {\bibfield  {journal} {\bibinfo  {journal}
  {Phys. Rev. Appl.}\ }\textbf {\bibinfo {volume} {13}},\ \bibinfo {pages}
  {064016} (\bibinfo {year} {2020})}\BibitemShut {NoStop}%
\bibitem [{\citenamefont {Palyanov}\ \emph {et~al.}(2016)\citenamefont
  {Palyanov}, \citenamefont {Kupriyanov}, \citenamefont {Borzdov},
  \citenamefont {Khokhryakov},\ and\ \citenamefont
  {Surovtsev}}]{palyanov2016high}%
  \BibitemOpen
  \bibfield  {author} {\bibinfo {author} {\bibfnamefont {Y.~N.}\ \bibnamefont
  {Palyanov}}, \bibinfo {author} {\bibfnamefont {I.~N.}\ \bibnamefont
  {Kupriyanov}}, \bibinfo {author} {\bibfnamefont {Y.~M.}\ \bibnamefont
  {Borzdov}}, \bibinfo {author} {\bibfnamefont {A.~F.}\ \bibnamefont
  {Khokhryakov}}, \ and\ \bibinfo {author} {\bibfnamefont {N.~V.}\ \bibnamefont
  {Surovtsev}},\ }\bibfield  {title} {\enquote {\bibinfo {title} {High-pressure
  synthesis and characterization of ge-doped single crystal diamond},}\ }\href
  {https://doi.org/10.1021/acs.cgd.6b00481} {\bibfield  {journal} {\bibinfo
  {journal} {Crystal Growth \& Design}\ }\textbf {\bibinfo {volume} {16}},\
  \bibinfo {pages} {3510--3518} (\bibinfo {year} {2016})}\BibitemShut {NoStop}%
\bibitem [{\citenamefont {Meesala}\ \emph {et~al.}(2018)\citenamefont
  {Meesala}, \citenamefont {Sohn}, \citenamefont {Pingault}, \citenamefont
  {Shao}, \citenamefont {Atikian}, \citenamefont {Holzgrafe}, \citenamefont
  {G\"undo\ifmmode~\breve{g}\else \u{g}\fi{}an}, \citenamefont {Stavrakas},
  \citenamefont {Sipahigil}, \citenamefont {Chia}, \citenamefont {Evans},
  \citenamefont {Burek}, \citenamefont {Zhang}, \citenamefont {Wu},
  \citenamefont {Pacheco}, \citenamefont {Abraham}, \citenamefont {Bielejec},
  \citenamefont {Lukin}, \citenamefont {Atat\"ure},\ and\ \citenamefont
  {Lon\ifmmode~\check{c}\else \v{c}\fi{}ar}}]{Strain2018Meesala}%
  \BibitemOpen
  \bibfield  {author} {\bibinfo {author} {\bibfnamefont {S.}~\bibnamefont
  {Meesala}}, \bibinfo {author} {\bibfnamefont {Y.-I.}\ \bibnamefont {Sohn}},
  \bibinfo {author} {\bibfnamefont {B.}~\bibnamefont {Pingault}}, \bibinfo
  {author} {\bibfnamefont {L.}~\bibnamefont {Shao}}, \bibinfo {author}
  {\bibfnamefont {H.~A.}\ \bibnamefont {Atikian}}, \bibinfo {author}
  {\bibfnamefont {J.}~\bibnamefont {Holzgrafe}}, \bibinfo {author}
  {\bibfnamefont {M.}~\bibnamefont {G\"undo\ifmmode~\breve{g}\else
  \u{g}\fi{}an}}, \bibinfo {author} {\bibfnamefont {C.}~\bibnamefont
  {Stavrakas}}, \bibinfo {author} {\bibfnamefont {A.}~\bibnamefont
  {Sipahigil}}, \bibinfo {author} {\bibfnamefont {C.}~\bibnamefont {Chia}},
  \bibinfo {author} {\bibfnamefont {R.}~\bibnamefont {Evans}}, \bibinfo
  {author} {\bibfnamefont {M.~J.}\ \bibnamefont {Burek}}, \bibinfo {author}
  {\bibfnamefont {M.}~\bibnamefont {Zhang}}, \bibinfo {author} {\bibfnamefont
  {L.}~\bibnamefont {Wu}}, \bibinfo {author} {\bibfnamefont {J.~L.}\
  \bibnamefont {Pacheco}}, \bibinfo {author} {\bibfnamefont {J.}~\bibnamefont
  {Abraham}}, \bibinfo {author} {\bibfnamefont {E.}~\bibnamefont {Bielejec}},
  \bibinfo {author} {\bibfnamefont {M.~D.}\ \bibnamefont {Lukin}}, \bibinfo
  {author} {\bibfnamefont {M.}~\bibnamefont {Atat\"ure}}, \ and\ \bibinfo
  {author} {\bibfnamefont {M.}~\bibnamefont {Lon\ifmmode~\check{c}\else
  \v{c}\fi{}ar}},\ }\bibfield  {title} {\enquote {\bibinfo {title} {Strain
  engineering of the silicon-vacancy center in diamond},}\ }\href {\doibase
  10.1103/PhysRevB.97.205444} {\bibfield  {journal} {\bibinfo  {journal} {Phys.
  Rev. B}\ }\textbf {\bibinfo {volume} {97}},\ \bibinfo {pages} {205444}
  (\bibinfo {year} {2018})}\BibitemShut {NoStop}%
\bibitem [{\citenamefont {Fehler}\ \emph {et~al.}(2021)\citenamefont {Fehler},
  \citenamefont {Antoniuk}, \citenamefont {Lettner}, \citenamefont {Ovvyan},
  \citenamefont {Waltrich}, \citenamefont {Gruhler}, \citenamefont {Davydov},
  \citenamefont {Agafonov}, \citenamefont {Pernice},\ and\ \citenamefont
  {Kubanek}}]{Fehler_2021}%
  \BibitemOpen
  \bibfield  {author} {\bibinfo {author} {\bibfnamefont {K.~G.}\ \bibnamefont
  {Fehler}}, \bibinfo {author} {\bibfnamefont {L.}~\bibnamefont {Antoniuk}},
  \bibinfo {author} {\bibfnamefont {N.}~\bibnamefont {Lettner}}, \bibinfo
  {author} {\bibfnamefont {A.~P.}\ \bibnamefont {Ovvyan}}, \bibinfo {author}
  {\bibfnamefont {R.}~\bibnamefont {Waltrich}}, \bibinfo {author}
  {\bibfnamefont {N.}~\bibnamefont {Gruhler}}, \bibinfo {author} {\bibfnamefont
  {V.~A.}\ \bibnamefont {Davydov}}, \bibinfo {author} {\bibfnamefont {V.~N.}\
  \bibnamefont {Agafonov}}, \bibinfo {author} {\bibfnamefont {W.~H.~P.}\
  \bibnamefont {Pernice}}, \ and\ \bibinfo {author} {\bibfnamefont
  {A.}~\bibnamefont {Kubanek}},\ }\bibfield  {title} {\enquote {\bibinfo
  {title} {Hybrid quantum photonics based on artificial atoms placed inside one
  hole of a photonic crystal cavity},}\ }\href
  {https://doi.org/10.1021/acsphotonics.1c00530} {\bibfield  {journal}
  {\bibinfo  {journal} {ACS Photonics}\ }\textbf {\bibinfo {volume} {8}},\
  \bibinfo {pages} {2635--2641} (\bibinfo {year} {2021})}\BibitemShut {NoStop}%
\bibitem [{\citenamefont {H{\"a}u{\ss}ler}\ \emph
  {et~al.}(2019{\natexlab{b}})\citenamefont {H{\"a}u{\ss}ler}, \citenamefont
  {Hartung}, \citenamefont {Fehler}, \citenamefont {Antoniuk}, \citenamefont
  {Kulikova}, \citenamefont {Davydov}, \citenamefont {Agafonov}, \citenamefont
  {Jelezko},\ and\ \citenamefont {Kubanek}}]{haussler2019preparing}%
  \BibitemOpen
  \bibfield  {author} {\bibinfo {author} {\bibfnamefont {S.}~\bibnamefont
  {H{\"a}u{\ss}ler}}, \bibinfo {author} {\bibfnamefont {L.}~\bibnamefont
  {Hartung}}, \bibinfo {author} {\bibfnamefont {K.~G.}\ \bibnamefont {Fehler}},
  \bibinfo {author} {\bibfnamefont {L.}~\bibnamefont {Antoniuk}}, \bibinfo
  {author} {\bibfnamefont {L.~F.}\ \bibnamefont {Kulikova}}, \bibinfo {author}
  {\bibfnamefont {V.~A.}\ \bibnamefont {Davydov}}, \bibinfo {author}
  {\bibfnamefont {V.~N.}\ \bibnamefont {Agafonov}}, \bibinfo {author}
  {\bibfnamefont {F.}~\bibnamefont {Jelezko}}, \ and\ \bibinfo {author}
  {\bibfnamefont {A.}~\bibnamefont {Kubanek}},\ }\bibfield  {title} {\enquote
  {\bibinfo {title} {Preparing single siv- center in nanodiamonds for external,
  optical coupling with access to all degrees of freedom},}\ }\href {\doibase
  10.1088/1367-2630/ab4cf7} {\bibfield  {journal} {\bibinfo  {journal} {New
  Journal of Physics}\ }\textbf {\bibinfo {volume} {21}},\ \bibinfo {pages}
  {103047} (\bibinfo {year} {2019}{\natexlab{b}})}\BibitemShut {NoStop}%
\bibitem [{\citenamefont {Albrecht}\ \emph {et~al.}(2013)\citenamefont
  {Albrecht}, \citenamefont {Bommer}, \citenamefont {Deutsch}, \citenamefont
  {Reichel},\ and\ \citenamefont {Becher}}]{albrecht2013coupling}%
  \BibitemOpen
  \bibfield  {author} {\bibinfo {author} {\bibfnamefont {R.}~\bibnamefont
  {Albrecht}}, \bibinfo {author} {\bibfnamefont {A.}~\bibnamefont {Bommer}},
  \bibinfo {author} {\bibfnamefont {C.}~\bibnamefont {Deutsch}}, \bibinfo
  {author} {\bibfnamefont {J.}~\bibnamefont {Reichel}}, \ and\ \bibinfo
  {author} {\bibfnamefont {C.}~\bibnamefont {Becher}},\ }\bibfield  {title}
  {\enquote {\bibinfo {title} {Coupling of a single nitrogen-vacancy center in
  diamond to a fiber-based microcavity},}\ }\href
  {https://doi.org/10.1103/PhysRevLett.110.243602} {\bibfield  {journal}
  {\bibinfo  {journal} {Physical review letters}\ }\textbf {\bibinfo {volume}
  {110}},\ \bibinfo {pages} {243602} (\bibinfo {year} {2013})}\BibitemShut
  {NoStop}%
\bibitem [{\citenamefont {Binder}\ \emph {et~al.}(2017)\citenamefont {Binder},
  \citenamefont {Stark}, \citenamefont {Tomek}, \citenamefont {Scheuer},
  \citenamefont {Frank}, \citenamefont {Jahnke}, \citenamefont {Müller},
  \citenamefont {Schmitt}, \citenamefont {Metsch}, \citenamefont {Unden},
  \citenamefont {Gehring}, \citenamefont {Huck}, \citenamefont {Andersen},
  \citenamefont {Rogers},\ and\ \citenamefont {Jelezko}}]{BINDER2017}%
  \BibitemOpen
  \bibfield  {author} {\bibinfo {author} {\bibfnamefont {J.~M.}\ \bibnamefont
  {Binder}}, \bibinfo {author} {\bibfnamefont {A.}~\bibnamefont {Stark}},
  \bibinfo {author} {\bibfnamefont {N.}~\bibnamefont {Tomek}}, \bibinfo
  {author} {\bibfnamefont {J.}~\bibnamefont {Scheuer}}, \bibinfo {author}
  {\bibfnamefont {F.}~\bibnamefont {Frank}}, \bibinfo {author} {\bibfnamefont
  {K.~D.}\ \bibnamefont {Jahnke}}, \bibinfo {author} {\bibfnamefont
  {C.}~\bibnamefont {Müller}}, \bibinfo {author} {\bibfnamefont
  {S.}~\bibnamefont {Schmitt}}, \bibinfo {author} {\bibfnamefont {M.~H.}\
  \bibnamefont {Metsch}}, \bibinfo {author} {\bibfnamefont {T.}~\bibnamefont
  {Unden}}, \bibinfo {author} {\bibfnamefont {T.}~\bibnamefont {Gehring}},
  \bibinfo {author} {\bibfnamefont {A.}~\bibnamefont {Huck}}, \bibinfo {author}
  {\bibfnamefont {U.~L.}\ \bibnamefont {Andersen}}, \bibinfo {author}
  {\bibfnamefont {L.~J.}\ \bibnamefont {Rogers}}, \ and\ \bibinfo {author}
  {\bibfnamefont {F.}~\bibnamefont {Jelezko}},\ }\bibfield  {title} {\enquote
  {\bibinfo {title} {Qudi: A modular python suite for experiment control and
  data processing},}\ }\href {\doibase
  https://doi.org/10.1016/j.softx.2017.02.001} {\bibfield  {journal} {\bibinfo
  {journal} {SoftwareX}\ }\textbf {\bibinfo {volume} {6}},\ \bibinfo {pages}
  {85--90} (\bibinfo {year} {2017})}\BibitemShut {NoStop}%
\end{thebibliography}%


\providecommand{\noopsort}[1]{}\providecommand{\singleletter}[1]{#1}%
\begin{thebibliography}{5}%
\makeatletter
\providecommand \@ifxundefined [1]{%
 \@ifx{#1\undefined}
}%
\providecommand \@ifnum [1]{%
 \ifnum #1\expandafter \@firstoftwo
 \else \expandafter \@secondoftwo
 \fi
}%
\providecommand \@ifx [1]{%
 \ifx #1\expandafter \@firstoftwo
 \else \expandafter \@secondoftwo
 \fi
}%
\providecommand \natexlab [1]{#1}%
\providecommand \enquote  [1]{``#1''}%
\providecommand \bibnamefont  [1]{#1}%
\providecommand \bibfnamefont [1]{#1}%
\providecommand \citenamefont [1]{#1}%
\providecommand \href@noop [0]{\@secondoftwo}%
\providecommand \href [0]{\begingroup \@sanitize@url \@href}%
\providecommand \@href[1]{\@@startlink{#1}\@@href}%
\providecommand \@@href[1]{\endgroup#1\@@endlink}%
\providecommand \@sanitize@url [0]{\catcode `\\12\catcode `\$12\catcode
  `\&12\catcode `\#12\catcode `\^12\catcode `\_12\catcode `\%12\relax}%
\providecommand \@@startlink[1]{}%
\providecommand \@@endlink[0]{}%
\providecommand \url  [0]{\begingroup\@sanitize@url \@url }%
\providecommand \@url [1]{\endgroup\@href {#1}{\urlprefix }}%
\providecommand \urlprefix  [0]{URL }%
\providecommand \Eprint [0]{\href }%
\providecommand \doibase [0]{http://dx.doi.org/}%
\providecommand \selectlanguage [0]{\@gobble}%
\providecommand \bibinfo  [0]{\@secondoftwo}%
\providecommand \bibfield  [0]{\@secondoftwo}%
\providecommand \translation [1]{[#1]}%
\providecommand \BibitemOpen [0]{}%
\providecommand \bibitemStop [0]{}%
\providecommand \bibitemNoStop [0]{.\EOS\space}%
\providecommand \EOS [0]{\spacefactor3000\relax}%
\providecommand \BibitemShut  [1]{\csname bibitem#1\endcsname}%
\let\auto@bib@innerbib\@empty
\bibitem [{\citenamefont {Fehler}\ \emph {et~al.}(2021)\citenamefont {Fehler},
  \citenamefont {Antoniuk}, \citenamefont {Lettner}, \citenamefont {Ovvyan},
  \citenamefont {Waltrich}, \citenamefont {Gruhler}, \citenamefont {Davydov},
  \citenamefont {Agafonov}, \citenamefont {Pernice},\ and\ \citenamefont
  {Kubanek}}]{Fehler_2021}%
  \BibitemOpen
  \bibfield  {author} {\bibinfo {author} {\bibfnamefont {K.~G.}\ \bibnamefont
  {Fehler}}, \bibinfo {author} {\bibfnamefont {L.}~\bibnamefont {Antoniuk}},
  \bibinfo {author} {\bibfnamefont {N.}~\bibnamefont {Lettner}}, \bibinfo
  {author} {\bibfnamefont {A.~P.}\ \bibnamefont {Ovvyan}}, \bibinfo {author}
  {\bibfnamefont {R.}~\bibnamefont {Waltrich}}, \bibinfo {author}
  {\bibfnamefont {N.}~\bibnamefont {Gruhler}}, \bibinfo {author} {\bibfnamefont
  {V.~A.}\ \bibnamefont {Davydov}}, \bibinfo {author} {\bibfnamefont {V.~N.}\
  \bibnamefont {Agafonov}}, \bibinfo {author} {\bibfnamefont {W.~H.~P.}\
  \bibnamefont {Pernice}}, \ and\ \bibinfo {author} {\bibfnamefont
  {A.}~\bibnamefont {Kubanek}},\ }\bibfield  {title} {\enquote {\bibinfo
  {title} {Hybrid quantum photonics based on artificial atoms placed inside one
  hole of a photonic crystal cavity},}\ }\href
  {https://doi.org/10.1021/acsphotonics.1c00530} {\bibfield  {journal}
  {\bibinfo  {journal} {ACS Photonics}\ }\textbf {\bibinfo {volume} {8}},\
  \bibinfo {pages} {2635--2641} (\bibinfo {year} {2021})}\BibitemShut {NoStop}%
\bibitem [{\citenamefont {Bayer}\ \emph {et~al.}(2022)\citenamefont {Bayer},
  \citenamefont {Berghaus}, \citenamefont {Sachero}, \citenamefont
  {Filipovski}, \citenamefont {Antoniuk}, \citenamefont {Lettner},
  \citenamefont {Waltrich}, \citenamefont {Klotz}, \citenamefont {Maier},
  \citenamefont {Agafonov},\ and\ \citenamefont {Kubanek}}]{bayer_2022}%
  \BibitemOpen
  \bibfield  {author} {\bibinfo {author} {\bibfnamefont {G.}~\bibnamefont
  {Bayer}}, \bibinfo {author} {\bibfnamefont {R.}~\bibnamefont {Berghaus}},
  \bibinfo {author} {\bibfnamefont {S.}~\bibnamefont {Sachero}}, \bibinfo
  {author} {\bibfnamefont {A.~B.}\ \bibnamefont {Filipovski}}, \bibinfo
  {author} {\bibfnamefont {L.}~\bibnamefont {Antoniuk}}, \bibinfo {author}
  {\bibfnamefont {N.}~\bibnamefont {Lettner}}, \bibinfo {author} {\bibfnamefont
  {R.}~\bibnamefont {Waltrich}}, \bibinfo {author} {\bibfnamefont
  {M.}~\bibnamefont {Klotz}}, \bibinfo {author} {\bibfnamefont
  {P.}~\bibnamefont {Maier}}, \bibinfo {author} {\bibfnamefont
  {V.}~\bibnamefont {Agafonov}}, \ and\ \bibinfo {author} {\bibfnamefont
  {A.}~\bibnamefont {Kubanek}},\ }\bibfield  {title} {\enquote {\bibinfo
  {title} {A quantum repeater platform based on single siv$^-$ centers in
  diamond with cavity-assisted, all-optical spin access and fast coherent
  driving},}\ }\href {https://arxiv.org/abs/2210.16157} {\bibfield  {journal}
  {\bibinfo  {journal} {arXiv}\ } (\bibinfo {year} {2022})}\BibitemShut
  {NoStop}%
\bibitem [{\citenamefont {Alkahtani}\ \emph {et~al.}(2019)\citenamefont
  {Alkahtani}, \citenamefont {Lang}, \citenamefont {Naydenov}, \citenamefont
  {Jelezko},\ and\ \citenamefont {Hemmer}}]{Alkahtani2019growth}%
  \BibitemOpen
  \bibfield  {author} {\bibinfo {author} {\bibfnamefont {M.}~\bibnamefont
  {Alkahtani}}, \bibinfo {author} {\bibfnamefont {J.}~\bibnamefont {Lang}},
  \bibinfo {author} {\bibfnamefont {B.}~\bibnamefont {Naydenov}}, \bibinfo
  {author} {\bibfnamefont {F.}~\bibnamefont {Jelezko}}, \ and\ \bibinfo
  {author} {\bibfnamefont {P.}~\bibnamefont {Hemmer}},\ }\bibfield  {title}
  {\enquote {\bibinfo {title} {Growth of high-purity low-strain fluorescent
  nanodiamonds},}\ }\href {\doibase 10.1021/acsphotonics.9b00224} {\bibfield
  {journal} {\bibinfo  {journal} {ACS Photonics}\ }\textbf {\bibinfo {volume}
  {6}},\ \bibinfo {pages} {1266--1271} (\bibinfo {year} {2019})},\ \Eprint
  {http://arxiv.org/abs/https://doi.org/10.1021/acsphotonics.9b00224}
  {https://doi.org/10.1021/acsphotonics.9b00224} \BibitemShut {NoStop}%
\bibitem [{\citenamefont {Xu}\ \emph {et~al.}(2023)\citenamefont {Xu},
  \citenamefont {Martin}, \citenamefont {Titze}, \citenamefont {Wang},
  \citenamefont {Sychev}, \citenamefont {Henshaw}, \citenamefont {Lagutchev},
  \citenamefont {Htoon}, \citenamefont {Bielejec}, \citenamefont {Bogdanov},
  \citenamefont {Shalaev},\ and\ \citenamefont
  {Boltasseva}}]{Xiaohui2023Fabrication}%
  \BibitemOpen
  \bibfield  {author} {\bibinfo {author} {\bibfnamefont {X.}~\bibnamefont
  {Xu}}, \bibinfo {author} {\bibfnamefont {Z.~O.}\ \bibnamefont {Martin}},
  \bibinfo {author} {\bibfnamefont {M.}~\bibnamefont {Titze}}, \bibinfo
  {author} {\bibfnamefont {Y.}~\bibnamefont {Wang}}, \bibinfo {author}
  {\bibfnamefont {D.}~\bibnamefont {Sychev}}, \bibinfo {author} {\bibfnamefont
  {J.}~\bibnamefont {Henshaw}}, \bibinfo {author} {\bibfnamefont {A.~S.}\
  \bibnamefont {Lagutchev}}, \bibinfo {author} {\bibfnamefont {H.}~\bibnamefont
  {Htoon}}, \bibinfo {author} {\bibfnamefont {E.~S.}\ \bibnamefont {Bielejec}},
  \bibinfo {author} {\bibfnamefont {S.~I.}\ \bibnamefont {Bogdanov}}, \bibinfo
  {author} {\bibfnamefont {V.~M.}\ \bibnamefont {Shalaev}}, \ and\ \bibinfo
  {author} {\bibfnamefont {A.}~\bibnamefont {Boltasseva}},\ }\bibfield  {title}
  {\enquote {\bibinfo {title} {Fabrication of single color centers in
  sub-50 nm nanodiamonds using ion implantation},}\ }\href {\doibase
  doi:10.1515/nanoph-2022-0678} {\bibfield  {journal} {\bibinfo  {journal}
  {Nanophotonics}\ }\textbf {\bibinfo {volume} {12}},\ \bibinfo {pages}
  {485--494} (\bibinfo {year} {2023})}\BibitemShut {NoStop}%
\bibitem [{\citenamefont {H{\"a}u{\ss}ler}\ \emph {et~al.}(2017)\citenamefont
  {H{\"a}u{\ss}ler}, \citenamefont {Thiering}, \citenamefont {Dietrich},
  \citenamefont {Waasem}, \citenamefont {Teraji}, \citenamefont {Isoya},
  \citenamefont {Iwasaki}, \citenamefont {Hatano}, \citenamefont {Jelezko},
  \citenamefont {Gali} \emph {et~al.}}]{haussler2017photoluminescence}%
  \BibitemOpen
  \bibfield  {author} {\bibinfo {author} {\bibfnamefont {S.}~\bibnamefont
  {H{\"a}u{\ss}ler}}, \bibinfo {author} {\bibfnamefont {G.}~\bibnamefont
  {Thiering}}, \bibinfo {author} {\bibfnamefont {A.}~\bibnamefont {Dietrich}},
  \bibinfo {author} {\bibfnamefont {N.}~\bibnamefont {Waasem}}, \bibinfo
  {author} {\bibfnamefont {T.}~\bibnamefont {Teraji}}, \bibinfo {author}
  {\bibfnamefont {J.}~\bibnamefont {Isoya}}, \bibinfo {author} {\bibfnamefont
  {T.}~\bibnamefont {Iwasaki}}, \bibinfo {author} {\bibfnamefont
  {M.}~\bibnamefont {Hatano}}, \bibinfo {author} {\bibfnamefont
  {F.}~\bibnamefont {Jelezko}}, \bibinfo {author} {\bibfnamefont
  {A.}~\bibnamefont {Gali}},  \emph {et~al.},\ }\bibfield  {title} {\enquote
  {\bibinfo {title} {Photoluminescence excitation spectroscopy of siv- and gev-
  color center in diamond},}\ }\href@noop {} {\bibfield  {journal} {\bibinfo
  {journal} {New Journal of Physics}\ }\textbf {\bibinfo {volume} {19}},\
  \bibinfo {pages} {063036} (\bibinfo {year} {2017})}\BibitemShut {NoStop}%
\end{thebibliography}%

\end{document}


\preprint{AIP/123-QED}

\title{ Supplementary Material - Enhanced Spectral Density of a Single Germanium Vacancy Center in a Nanodiamond by Cavity-Integration}   

\author{Florian Feuchtmayr}
\altaffiliation{F.F. and R.B.  contributed equally to this work.}
\affiliation{Institute for Quantum Optics, Ulm University, Albert-Einstein-Allee 11, 89081 Ulm, Germany} 
\author{Robert Berghaus}
\altaffiliation{F.F. and R.B.  contributed equally to this work.}
\affiliation{Institute for Quantum Optics, Ulm University, Albert-Einstein-Allee 11, 89081 Ulm, Germany}
\author{Selene Sachero}
\affiliation{Institute for Quantum Optics, Ulm University, Albert-Einstein-Allee 11, 89081 Ulm, Germany} 
\author{Gregor Bayer}
\affiliation{Institute for Quantum Optics, Ulm University, Albert-Einstein-Allee 11, 89081 Ulm, Germany} 
\author{Niklas Lettner}
\affiliation{Institute for Quantum Optics, Ulm University, Albert-Einstein-Allee 11, 89081 Ulm, Germany} 
\affiliation{Center for Integrated Quantum Science and Technology (IQst), Ulm University, Albert-Einstein-Allee 11, 89081 Ulm, Germany}
\author{Richard Waltrich}
\affiliation{Institute for Quantum Optics, Ulm University, Albert-Einstein-Allee 11, 89081 Ulm, Germany} 
\author{Patrick Maier}
\affiliation{Institute for Quantum Optics, Ulm University, Albert-Einstein-Allee 11, 89081 Ulm, Germany} 
\author{Viatcheslav Agafonov}
\affiliation{GREMAN, UMR 7347 CNRS, INSA-CVL, Tours University, 37200 Tours, France}
\author{Alexander Kubanek}
\altaffiliation[Corresponding author: ]{alexander.kubanek@uni-ulm.de}
\affiliation{Institute for Quantum Optics, Ulm University, Albert-Einstein-Allee 11, 89081 Ulm, Germany}
\affiliation{Center for Integrated Quantum Science and Technology (IQst), Ulm University, Albert-Einstein-Allee 11, 89081 Ulm, Germany}

\date{\today}

\maketitle

\renewcommand\thefigure{S\arabic{figure}} 
\renewcommand\theequation{S\arabic{equation}}
\renewcommand\thetable{S\arabic{table}}

\section{ Debye-Waller factor}
\begin{figure}[h]
    \centering
    \includegraphics[width=0.68\linewidth]{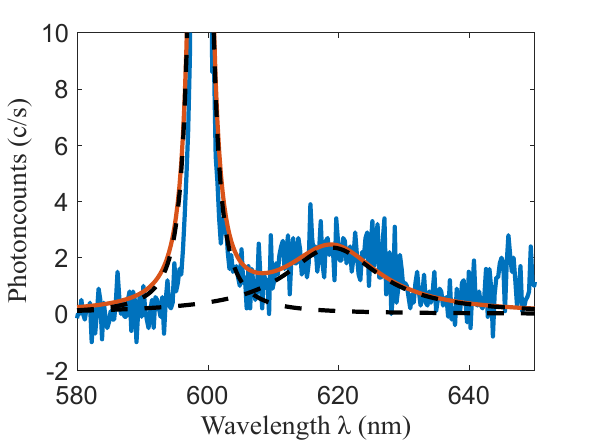}
    \caption{ Magnified PL spectrum of fig. 1(c) of the main manuscript with a coarse grating of 150 g/mm of ND GeV-I. The ZPL and PSB are fitted with Lorentzian functions. Orange fit shows the sum of the two functions.}
    \label{fig4}
\end{figure}

In order to determine the DW factor, the counts of the ZPL and the counts of the PSB were added up separately up to a wavelength of 700\,nm. Comparing these counts with
\begin{equation}
    \text{DW}=\frac{\text{ZPL}}{\text{ZPL}+\text{PSB}}
\end{equation}
yielded a DW factor of 0.6. On the other hand, the DW factor was calculated to 0.7 by integrating the fits of the ZPL and PSB as seen in fig. \ref{fig4}. Therefore, 0.6 is given in the main manuscript as a lower bound.

\section{Autocorrelation Measurements}

\begin{figure}[h]
    \centering
    \includegraphics[width=\linewidth]{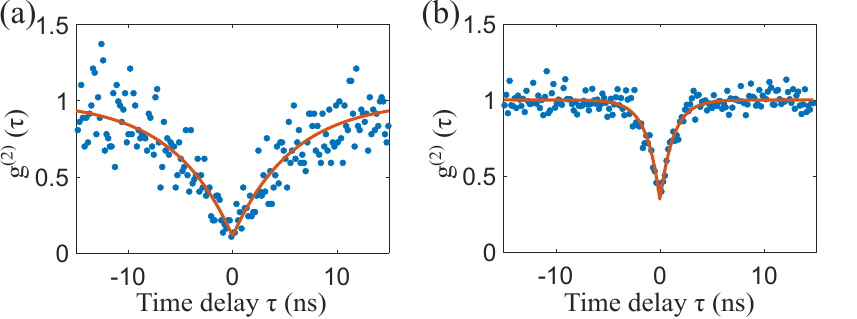}
    \caption{Off-resonant autocorrelation measurements without background correction. (a) ND GeV-II showing anti-bunching of $g^{(2)}(0)$=0.11\,$\pm$\,0.07 and an excited state lifetime of $\tau_{\textup{LT}}$=(5.8\,$\pm$\,0.9)\,ns at (75\,$\pm$\,3)\,$\upmu$W. (b) ND GeV-III showing anti-bunching of $g^{(2)}(0)$=0.34\,$\pm$\,0.03 and an excited state lifetime of $\tau_{\textup{LT}}$=(1.28\,$\pm$\,0.09)\,ns at (42\,$\pm$\,2)\,$\upmu$W.}
    \label{fig5}
\end{figure}

The sample from which ND GeV-I was taken also contained further NDs with single defects e.g. ND GeV-II and ND GeV-III. Their autocorrelation measurements under off-resonant excitation are depicted in figure \ref{fig5} and show antibunching. 
It can be concluded that the emission of these NDs is dominated by single GeV$^-$ centers with lifetimes ranging between $\tau=1.3\,-\,5.8$\,ns.

\section{ Nanomanipulation of Nanodiamonds}

 The employed pick and place technique used here is comparable to earlier works: Fehler et al.
 \cite{Fehler_2021} 
 transfer of a SiV-ND to a photonic crystal cavity) and Bayer et al.
\cite{bayer_2022} 
transfer of a SiV-ND to a focused ion beam milled cavity structure).

To scan and identify the NDs before the transfer we used a platinum cantilever (tip: HQ:NSC15\slash Pt).
The selected ND could be picked up using contact mode. Having the ND attached to the cantilever and the upper mirror mount with the CO$_2$ fabricated cavity structures mounted inside the AFM the structures could be located by scanning in contactless tapping mode. 
From this imaging the center of the cavity structure could be identified, where the ND was then placed.
For a more precise (AFM resolution) positioning of the ND contact nanomanipulation was used to push the ND with a low-adhesion cantilever (tip: 160AC-NA). AFM nanomanipulation of the ND was performed with the help of the JPK nanomanipulation toolbox.

With a ND centered in the cavity mirror we then align our optical cavity on a maximized TEM00 mode in order to overlap the positioned ND with the cavity field.

The presented universal cavity integration can be carried out for other ND and other nanoparticles from different fabrication methods. 
Advancements in the production methods, such as employing ND growth under lower growth temperatures,\cite{Alkahtani2019growth}
or ion implantation
\cite{Xiaohui2023Fabrication}
could enable small (sub- 50\,nm) NDs  containing single, spectral stable  GeV$^-$ center in the future. Therewith, the finesse could be increased significantly.

\section{Spectral Enhancement Calculations}
\begin{table}[htb]
    \centering 
    \caption{SDE calculations of ND GeV-I including different setup efficiencies. For better comparison the enhancements are calculated in respect to the free space emitter.}
    \label{tab1}
    \begin{tabular}{| l | rrr | }
     \hline     
    \textbf{Setup} &  \textbf{Confocal 1} & \textbf{Confocal 2} & \textbf{Cavity} \\ 
    \hline     \hline
Photoncounts in c/s  & 1,956\,$\pm$\,40 & 234\,$\pm$\,10 & 72\,$\pm$\,3 \\ 
FWHM in GHz & 1,036\,$\pm$\,25 & 918\,$\pm$\,25 & 1.9\,$\pm$\,0.5 \\  
\textit{P} in $\upmu$W   & 930\,$\pm$\,20 & 20\,$\pm$\,1 & 80\,$\pm$\,3 \\ 
\hline 
\textbf{SD} & \multirow{2}{*}{2.03\,$\pm$\,0.08} & \multirow{2}{*}{12.7\,$\pm$\,1.0} & \multirow{2}{*}{470\,$\pm$\,120}   \\
in c/(s GHz mW) &  &  &    \\
     \hline      \hline
Exc. efficiency ($10^{-5}$) & 22\,$\pm$\,2 & 25\,$\pm$\,1 & 0.07\,$\pm$\,0.03 \\  
Spectral efficiency & 1 (532\,nm) & 1 (532\,nm) & 2 (587\,nm)\\ 
Weighting of finesse & 1  & 2 & 1,800\,$\pm$\,800\\ 
     \hline
$\eta^{\textup{exc}}$ ($10^{-4}$) & 2.2\,$\pm$\,0.2 & 5.0\,$\pm$\,0.2 & 23\,$\pm$\,7 \\  
     \hline     \hline 
$\eta^{\textup{det}}$ ($10^{-2}$) & 1.2\,$\pm$\,0.1 & 1.5\,$\pm$\,0.2 & 0.52\,$\pm$\,0.05 \\ 
     
     \hline     \hline
\textbf{SDE} & 1 & 2.1\,$\pm$\,0.4 & 48\,$\pm$\,20  \\
     \hline     
     \end{tabular}
\end{table}

We analyzed the spectra of ND GeV-I in figures 1(d), 2(f) and 3(d)  of the main manuscript, which were acquired in confocal microscope setups I, II and the cavity setup, respectively. The SDs of each spectrum were extracted and the total excitation and detection efficiencies ($\eta^{\textup{exc}}$, $\eta^{\textup{det}}$) of each setup were determined, see table \ref{tab1}. From these values, the SDE with respect to Confocal I is given by 
\begin{equation}
    \textup{SDE}=\frac{\textup{SD}}{\textup{SD}_{\textup{Con1}}}\times\frac{\eta_{\textup{Con1}}^{\textup{det}}}{\eta^{\textup{det}}}\times\frac{\eta_{\textup{Con1}}^{\textup{exc}}}{\eta^{\textup{exc}}}.
\end{equation}
For the SD, the FWHM at the confocal configurations were obtained by the Lorentzian fit of the ZPL. The FWHM of the cavity configuration was given by FWHM\,=\,FSR\,/\,$\mathcal{F}$ with FSR\,=\,c\,/\,2L and $\mathcal{F}$ at the emitter.\\ 
The total excitation efficiency includes the ratio of the emitter area and the beam area, the relative efficiency of different excitation wavelengths \cite{haussler2017photoluminescence} and the effective excitation power by weighting the finesse. The detection efficiency takes the transmission of every optical component of the three different setups into account. 


\section*{References}

\bibliography{si}